\documentclass[prx,twocolumn]{revtex4}  

\usepackage{amssymb}
\usepackage{amsmath}
\usepackage{graphicx}
\usepackage{bbm}
\usepackage{color}
\usepackage{comment}
\usepackage{gensymb}
\usepackage{gensymb}
\usepackage{cancel} 
\usepackage{array}
\usepackage[x11names]{xcolor} 

\newcommand{\fullcell}[3]{%
  \colorbox{#1}{\parbox[c][10pt][c]{#2}{\centering \arraybackslash \large#3}}%
}

\definecolor{yellow}{RGB}{255, 255, 153}
\definecolor{blue}{RGB}{146,205,251}
\definecolor{gray}{RGB}{176,177,177}
\definecolor{green}{RGB}{188,233,161}

\newcommand{\ia}[1]{\textcolor{black}{#1}}

\begin{document}

\title{Engineering protected cavity-QED interactions through pulsed dynamical decoupling}

\author{I. Arrazola$^{1,2}$, P. Bertet$^{3}$, Y. Chu$^{4}$,  P. Rabl$^{5,6,7}$}
\affiliation{$^1$EHU Quantum Center and Department of Physics, University of the Basque Country UPV/EHU, 48080 Bilbao, Spain}
\affiliation{$^2$Instituto de F\'isica Te\'orica, UAM-CSIC, Universidad Aut\'onoma de Madrid, Cantoblanco, 28049 Madrid, Spain}
\affiliation{$^3$Quantronics Group, SPEC, CEA Saclay, CNRS, Université Paris-Saclay, 91191 Gif-sur-Yvette, France}
\affiliation{$^4$Department of Physics, ETH Z\"urich,  8093 Zurich, Switzerland}
\affiliation{$^5$Walther-Mei\ss ner-Institut, Bayerische Akademie der Wissenschaften, 85748 Garching, Germany}
\affiliation{$^6$Technische Universit\"at M\"unchen, TUM School of Natural Sciences, Physics Department, 85748 Garching, Germany} 
\affiliation{$^7$Munich Center for Quantum Science and Technology (MCQST), 80799 Munich, Germany} 

\date{\today}

\begin{abstract}
We study a generic cavity QED setup under conditions where the coupling between the two-level systems and a single bosonic mode is significantly degraded by low-frequency noise. To overcome this problem, we identify pulsed dynamical decoupling strategies that suppress the effects of noise while still allowing for a coherent exchange of excitations between the individual subsystems. The corresponding pulse sequences can be further designed to realize either Jaynes-Cummings, anti-Jaynes-Cummings, or Rabi couplings, as well as different types of cavity-mediated interactions between the two-level systems. A detailed analysis of the residual imperfections demonstrates that this decoupling strategy can boost the effective cooperativity of the cavity QED system by several orders of magnitude and improve the fidelity of quantum-technologically relevant operations accordingly.   
\end{abstract}

\maketitle

\section{Introduction}

The Jaynes-Cummings (JC) model describes the near-resonant coupling of a two-level system (TLS) to a single bosonic mode and for many decades it served as a prototypical toy model for studying light-matter interactions at the quantum level~\cite{Jaynes:1963,HarocheBook,JCM_Book}. In recent years, the JC model has regained considerable interest in the context of quantum technologies, where it describes the basic processes relevant for generating non-classical photonic states~\cite{Law:1996} or for realizing qubit-photon interfaces~\cite{Cirac:1997}. The cavity mode can also be used to implement long-range interactions between two or more TLSs, as it has already been demonstrated in a variety of systems ranging from optical cavity QED~\cite{Welte:2018} and trapped ions~\cite{SchmidtKaler:2003} to superconducting circuits~\cite{Maier:2007} and solid-state spin qubits~\cite{Dijkema:2025}. For all of these applications it is required that the coupling between the TLS and the bosonic mode is sufficiently strong in order to overcome the relevant decoherence processes in the system.

While in many of the originally considered cavity QED experiments with atoms and optical photons the decoherence rate of the TLS is mainly determined by the decay rate of the excited atomic state, this is not necessarily the case in other experimental platforms, where equivalent interactions are studied. Prominent examples include optical cavity QED systems with rare-earth dopants~\cite{Zhong:2019,Kindem:2020,Ourari:2023,Gritsch:2025} as well as spin ensembles~\cite{Schuster:2010,Kubo:2010,Amsuss:2011,Probst:2013}, individual impurity spins~\cite{Wang2023} and gate-defined quantum dots~\cite{Dijkema:2025,Mi:2018,Samkharadze:2018,Landig:2018} coupled to microwave resonators. In those and many other systems of interest, spontaneous decay is often negligible compared to the dephasing rate $\Gamma_\phi=1/T_2^*$ associated with inhomogeneous broadening, low-frequency magnetic noise or other slow parameter drifts. For isolated TLSs, it is well-known that quasi-static shifts can be effectively suppressed using spin-echo techniques or more advanced pulsed dynamical decoupling (DD) schemes~\cite{Biercuk:2009,Bluhm:2011,Bylander:2011,Peng:2011,Wang:2016}.  However, applying pulsed DD in cavity QED systems presents unique challenges, as fast $\pi$-rotations also disrupt or cancel~\cite{Pryadko:2008} the coherent interaction between the TLS and the cavity mode. This hinders a straightforward adaption of DD techniques in such systems.

A possible way to overcome this problem is to continuously drive the TLS with a strong external field.  The fast Rabi oscillations then average out any quasi-static energy shifts~\cite{Koppens:2006,Yan:2013,Golter:2014,Barfuss:2015,Laucht:2017,Miao:2020}, while still permitting a resonant interaction between the cavity mode and the resulting dressed qubit states~\cite{Rabl:2009}, or the realization of protected quantum gates~\cite{Timoney:2011,Tan:2013,Guo:2018,Cao:2017,Srinivasa:2024,Arrazola:2024}. However, this continuous DD technique comes with several practical limitations. In particular, it is often difficult to control the power of the driving field with sufficient accuracy~\ia{\cite{Cai:2012,Teissier:2017}} and the application of a strong and continuous driving field can lead to undesired heating effects~\ia{\cite{Krinner:2019}}. \ia{With few exceptions \cite{Guo:2018,Cornell2025} these complications have hindered a widespread use of continuous DD schemes so far, especially in cryogenic experimental setups.}

In this paper, we present an alternative approach for protecting cavity QED systems against low-frequency noise, leveraging the well-established pulsed DD techniques for individual TLSs~\cite{Slichter:Book,Viola:1998,Ezzell:2023}. \ia{In particular, going beyond a specific sequence that has been previously proposed for this problem~\cite{Groszkowski:2022}, we present a general framework for identifying and characterizing sequences of fast $\pi$-rotations,} which recover the JC model as an effective interaction, despite the fact that the system dynamics is repeatedly interrupted. Further, we find that different variants of these pulse sequences can be used to engineer effective anti-JC or Rabi-type interactions, which are not present in the original system. The same approach can also be applied to protect cavity-mediated interactions between two or more TLSs, with a similar flexibility in the design of the effective interactions through an appropriate choice of pulse parameters.

From a detailed analysis of these processes, we find that noise-induced errors can be systematically suppressed by increasing the number of applied $\pi$-pulses, $N_\pi$. Specifically, for cavity-mediated quantum gates, the residual error scales as $ \mathcal{E}\sim 1/(\mathcal{C}N_{\rm \pi})^{4/5}$, where $\mathcal{C}$ is the cooperativity of the bare cavity QED system. This almost linear improvement can be used to substantially boost the fidelity of cavity-mediated gate operations in existing setups, but also to enable the experimental realization of new cavity QED platforms that have so far been hindered by the presence of excessive noise. More generally, the techniques described in this paper extend previous schemes for pulsed Hamiltonian engineering for interacting spin systems~\cite{Sollsteimer:2001,Hayes:2014,Choi:2020,Morong:2023,Zhou:2024} to a more general set of spin-boson-type models, as relevant in quantum optics and various areas of solid-state physics.

\begin{figure}
	\includegraphics[width=\columnwidth]{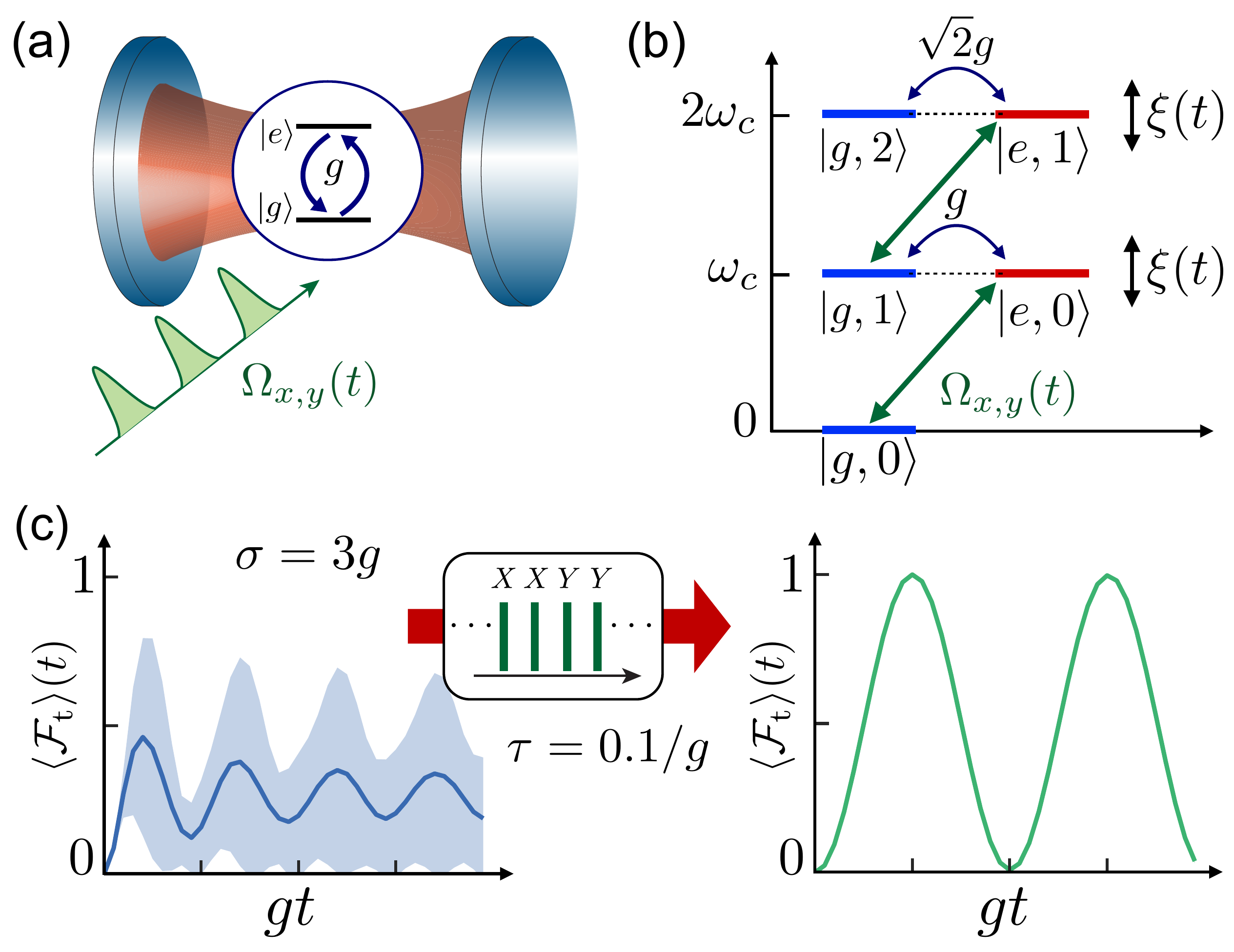}
	\caption{(a) Sketch of a cavity QED setup, where a TLS is coupled to a near-resonant bosonic mode. (b) Energy diagram of the relevant states of the JC model. In the presence of noise, the states $|e,n\rangle$ are shifted compared to the states $|g,n\rangle$ by a slowly fluctuating frequency $\xi(t)$, which leads to dephasing of the bare TLS with a rate $1/T_2^*$. (c) Simulation of the vacuum Rabi oscillations for the case where the dephasing rate is comparable to the coupling strength (left panel). Under the same conditions, but interrupting this evolution by an appropriate sequence of $\pi$-pulses, the effect of the noise can be significantly suppressed while preserving the coherent oscillations between the TLS and the cavity mode (right panel). See text for more details.}  
	\label{Fig:1_Setup}
\end{figure}

\section{The noisy Jaynes-Cummings model}
 We consider a generic cavity QED setup as shown in Fig.~\ref{Fig:1_Setup}(a), where a single TLS is coupled to a near resonant bosonic mode with frequency $\omega_c$ and annihilation and creation operators $a$ and $a^\dag$. The ground state $|g\rangle$ and the excited state $|e\rangle$ of the TLS are separated by a  bare transition frequency $\omega_0$. In a rotating frame with respect to this frequency and under the validity of the rotating-wave approximation, the system is described by the JC Hamiltonian $(\hbar=1)$
\begin{equation}\label{eq:JCModel}
H_{\rm JC}(t)=   \Delta  a^\dagger a+ g(\sigma_+ a+\sigma_-a^\dagger ) +\frac{\xi(t)}{2}\sigma_z ,
\end{equation}
where  $g$  is the coupling strength and $\Delta=\omega_c-\omega_0$ is the detuning. In Eq.~\eqref{eq:JCModel}, $\xi(t)$ accounts for an additional unknown frequency shift of the TLS, which describes, for example, the effect of magnetic field fluctuations or other sources of low-frequency noise.  For concreteness, we model $\xi(t)$ in terms of an Ornstein-Uhlenbeck process~\cite{Gillespie96} with zero mean, $\langle \xi(t)\rangle=0$, and 
\begin{equation}
\langle \xi(t)\xi (0)\rangle= \sigma^2e^{-t/\tau_c}.
\end{equation}
Here $\sigma$ and $\tau_c$ quantify the strength of the noise and its correlation time, respectively. 

In this work we are primarily interested in the experimentally relevant regime of slow noise, $\tau_c \gg \tau_\pi$, in which case fast DD pulses with a duration $\tau_\pi$ can be applied to suppress the low-frequency components of the noise. For this purpose, we consider a total Hamiltonian of the form
\begin{equation}\label{eq:totalHamiltonian}
H(t)=H_{\rm JC}(t) + H_{\rm drive}(t),
\end{equation}
where
\begin{equation}
H_{\rm drive}(t)=\frac{\Omega_x(t)}{2}\sigma_x + \frac{\Omega_y(t)}{2}\sigma_y
\end{equation}
accounts for external driving fields with switchable Rabi frequencies $\Omega_x(t)$ and $\Omega_y(t)$. These control fields are used to implement fast $\pi$-rotations of the TLS along the $x$-axis (``$X$ pulse") or the $y$-axis (``$Y$ pulse"), respectively.

For concreteness, in the following analysis we assume that the driving fields can be well-approximated by rectangular pulses, \ia{which are standard in experiments and provide the fastest $\pi$-pulses. These are described by}
\begin{equation}\label{eq:OmegaPulses}
\Omega_{\eta=x,y}(t)= \sum_i  \left(\pi + \delta \theta_i^{\eta} \right) \Theta(t-t^{\eta}_i).
\end{equation}
Here the $t^{x}_i$ ($t^{y}_i$) denote the times at which the individual $X$ ($Y$) pulses are applied and the window function  assumes a value of $\Theta(t)=1/\tau_\pi$ for $t\in\{-\tau_\pi/2,\tau_\pi/2\}$ and $\Theta(t)=0$ otherwise. Therefore, for $\delta \theta_i^{x,y}=0$, the driving field implements a series of complete $\pi$-rotations of the form $X= e^{-i \sigma_x\pi/2}$ and $Y= e^{-i \sigma_y\pi/2}$. Finite values of \ia{$|\delta \theta_{i}^{x,y}|\ll \pi$ correspond to small under- or over-rotations that may arise from pulse imperfections. In addition, such pulse deviations can also be deliberately introduced in a deterministic manner to engineer specific effective interactions, as explained below}.

\subsection{Vacuum Rabi oscillations vs spin echo}\label{subsec:vaccumRabi}
In the absence of noise, the JC Hamiltonian $H_{\rm JC}$ preserves the total number of excitations and induces coherent oscillations between the states $|e,n-1\rangle$ and $|g,n\rangle$ with a photon-number dependent Rabi-frequency $2g\sqrt{n}$ on resonance [see Fig.~\ref{Fig:1_Setup}(b)]. Specifically, by initializing the system in state $|\psi(0)\rangle=|e,0\rangle$, the state evolves as 
\begin{equation}\label{eq:DetJCEvolution}
|\psi(t)\rangle=  C(t)|e,0\rangle -i S(t)|g,1\rangle,
\end{equation} 
where $C(t)=\cos(\tilde{g} t)-i \sin(\tilde{g} t)\cos (\theta)$ and $S(t)= \sin(\tilde{g} t)\sin (\theta)$, with $\tilde g= \sqrt{g^2+\Delta^2/4}$ and a mixing angle given by $\tan(\theta)=2g/\Delta$. For $\Delta=0$ ($\theta=\pi/2$) and a time $T_{\rm t}=\pi/(2g)$, the excitation is completely transferred from the TLS to the cavity mode, i.e. $|\psi(T_{\rm t})\rangle=|g,1\rangle$. Therefore, this process serves as a basic ingredient for preparing the cavity in a non-classical state or for realizing a qubit-photon interface.

This transfer quickly deteriorates, once random frequency fluctuations with $\sigma \gtrsim g$ are introduced. For static noise, this effect can be understood from Eq.~\eqref{eq:DetJCEvolution}, by replacing the known detuning $\Delta$ by a random frequency offset $\xi$. Fig.~\ref{Fig:1_Setup}(c) shows the resulting averaged state-transfer fidelity, $\langle\mathcal{F}_{\rm t} (t)\rangle$, along with its standard deviation (indicated by the shaded area). Here, $\mathcal{F}_{\rm t} (t)=|\langle g,1|\psi(t)\rangle|^2$ and $\langle \cdot \rangle$ denotes the average over different noise realizations.  For small noise strength, 
$\sigma/g\ll 1$, the fidelity decays as (see Appendix~\ref{app:TranferProb})
 \begin{equation}\label{eq:bare_fidelity}
 \langle \mathcal{F}_{\rm t}(t=T_{\rm t})\rangle \approx 1-\left(\frac{\sigma}{2g}\right)^2.
 \end{equation} 
This result confirms our basic intuition that the coupling strength $g$ must considerably exceed the noise strength $\sigma$, in order to permit a coherent exchange of excitations. In many experimental settings, this regime cannot be reached\ia{~\cite{Probst:2013}}.

For the case of an uncoupled TLS, the effect of low-frequency noise can be efficiently suppressed by applying a fast $\pi$-rotation during half of the evolution. However, as can be seen from Eq.~\eqref{eq:DetJCEvolution} and Fig.~\ref{Fig:1_Setup}(b), flipping the states $|g\rangle$ and $|e\rangle$ at a time $T_{\rm t}/2$ would populate the states $|g,0\rangle$ and $|e,1\rangle$, which live in different excitation-number subspaces and are decoupled from the target state $|g,1\rangle$ in the successive evolution. Therefore, while such a simple decoupling approach would reduce the impact of the noise, it would also completely spoil the state transfer.

\section{Dynamically protected vacuum Rabi oscillations}\label{sec:ProtectedJC}
 Let us now address how we can use DD pulses to suppress noise while preserving the coherent Rabi oscillations between the TLS and the cavity mode.  To do so, we consider the dynamics of the noisy JC model, interrupted by a series of fast $\pi$-rotations. To identify an appropriate decoupling strategy, we change to a so-called toggling frame~\cite{Brinkmann:2016} via the unitary transformation 
\begin{equation}
|\psi(t)\rangle= U_{\pi}(t)e^{-i\Delta t \,a^\dag a }|\tilde \psi(t)\rangle.
\end{equation} 
Here, $U_{\pi}(t) = \mathcal{T}e^{-i \int_0^t ds H_{\rm drive}(s)}|_{\delta \theta_i^{x,y}=0}$ describes the bare evolution of the TLS under the influence of the external driving field, assuming perfect $\pi$-pulses. In this new frame, the system evolves according to the Hamiltonian
\begin{equation}\label{eq:TogglinFrameHamil}
\tilde H(t)= \tilde H_{\rm int}(t) +\delta \tilde H_{\rm drive}(t) +
\frac{\xi(t)}{2}\tilde\sigma_z(t),
\end{equation} 
where we have defined $\tilde O(t)= U_{\pi}^\dag(t) O U_{\pi}(t)$ and 
 \begin{equation}
\tilde H_{\rm int}(t) = g\left[\tilde \sigma_-(t) a^\dagger  e^{i\Delta t} +{\rm H.c.} \right]
\end{equation} 
is the interaction Hamiltonian in the toggling frame. In Eq.~\eqref{eq:TogglinFrameHamil}, we have further included the term 
 \begin{equation}
\delta \tilde H_{\rm drive}(t)= \frac{\delta\Omega_x(t)}{2}\tilde \sigma_x(t) + \frac{\delta\Omega_y(t)}{2}\tilde \sigma_y(t)
\end{equation} 
to account for deviations from the complete $\pi$-rotations assumed in the definition of $U_\pi$. Note that according to Eq.~\eqref{eq:OmegaPulses}, $\delta\Omega_{x,y}(t)= \sum_i   \delta \theta_i^{x,y} \Theta(t-t^{x,y}_i)$.

\begin{figure}
\includegraphics[width=\columnwidth]{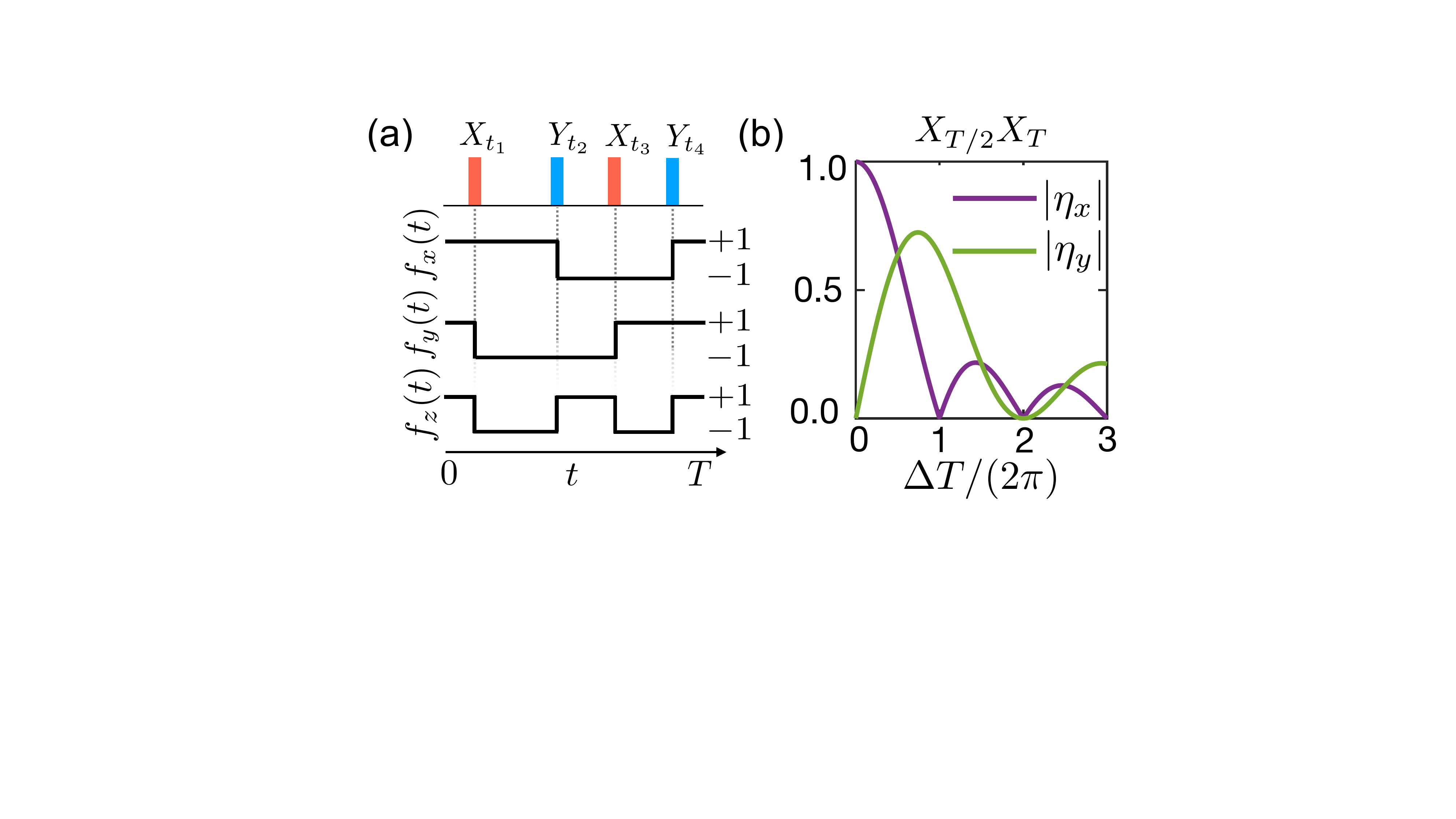}
	\caption{(a) Plots of the modulation functions $f_{x,y,z}(t)$ for an example pulse sequence with four $\pi$-rotations (along both $x$ and $y$ axes) applied within a period $T$. (b) Dependence of the absolute values of $\eta_{x,y}$ on the detuning $\Delta$ for the basic sequence $X_{T/2}X_T$.}
	\label{Fig:2}
\end{figure}

\subsection{Interaction engineering} 
While in all our numerical simulations a finite duration of the $\pi$-pulses is taken into account, for the following analytic considerations we restrict ourselves to instantaneous $\pi$-rotations and denote a specific pulse sequence by $X_{t^x_1} Y_{t_1^y} Y_{t_2^y}X_{t_2^x}\dots $, etc. In this limit we find that
 \begin{equation}\label{eq:fliprules}
U_{\pi}^\dag(t)  \sigma_{k} U_{\pi} (t) =  f_k(t)\sigma_k,\qquad k=x,y,z,
\end{equation} 
where $f_k(t)=\pm 1$. Specifically, as illustrated in Fig.~\ref{Fig:2}(a), the function $f_x(t)$ changes sign whenever a $Y$ rotation is applied (i.e. at times $t_i^y$) and $f_y(t)$ changes sign whenever an $X$ rotation is applied (i.e. at times $t_i^x$). For the $z$ component we obtain $f_z(t)=f_x(t) f_y(t)$, which changes its sign after every pulse.

Given a pulse sequence with a total duration $T$ that is short compared to $g^{-1}$, we can use first-order perturbation theory to approximate the system evolution after this interval by a time-averaged effective Hamiltonian~\cite{Brinkmann:2016}
\begin{equation}
\tilde H_{\rm eff} = \frac{1}{T} \int_{0}^{T} ds \, \tilde H(s) = \tilde H^{\rm eff}_{\rm int} + \delta \tilde H^{\rm eff}_{\rm drive} +\tilde H^{\rm eff}_{\rm noise}.
\end{equation} 
Here, the first term,
\begin{equation}\label{eq:effHamil}
\tilde H^{\rm eff}_{\rm int} = \frac{g}{2}\left[(\eta_x  \sigma_x - i \eta_y \sigma_y)   a^\dagger  +{\rm H.c.} \right],
\end{equation} 
is the effective interaction Hamiltonian with complex parameters
 \begin{eqnarray}\label{eq:etaxy}
\eta_{x,y} & = & \frac{1}{T} \int_{0}^{T} ds f_{x,y}(s) e^{i\Delta s}.
\end{eqnarray}
This expression can be used to identify an appropriate set of decoupling pulses to obtain the desired effective light-matter coupling for this period. In Fig.~\ref{Fig:2}(b), the values of $|\eta_{x,y}|$ are plotted as a function of the detuning $\Delta$ for a simple DD sequence $X_{T/2}X_{T}$.  As one can already see from this plot, the form of the effective interaction depends crucially on the value of $\Delta$. For example, for $\Delta=0$, the interaction will be of the form $\tilde{H}_{\rm int}^{\rm eff} \sim \sigma_x (a + a^\dag)$, while for $\Delta = 4\pi/T$, the interaction cancels out completely.

To extend this effective interaction to arbitrary times, we consider periodic pulse sequences with $f_{k}(t+T)=f_{k}(t)$. Then, the effective parameters $\eta^{(n)}_{x,y}$ for the $n$-th time interval $\{n T,(n+1)T\}$, fulfill
 \begin{eqnarray}\label{eq:etan}
\eta_{x,y}^{(n)} &=&e^{i n \Delta T} \eta_{x,y}.
\end{eqnarray}
By setting $\Delta T= 2\pi m+ \Delta_{\rm eff}  T$, where $m\in \mathbb{Z}$ and $\Delta_{\rm eff} $ is a residual detuning, we can rewrite the effective interaction Hamiltonian for all time intervals as
\begin{equation}\label{eq:effHamil2}
\tilde H_{\rm int}^{\rm eff}(nT) = \frac{g_{\rm eff}}{2} \left(\vec{n}\cdot\vec{\sigma} \,a^\dagger e^{i\Delta_{\rm eff} nT}  +{\rm H.c.} \right).
\end{equation} 
Here $g_{\rm eff}=\eta g$ is the effective coupling strength, $\vec{n}=(\eta_x,-i\eta_y,0)/\eta$ and $\eta=(|\eta_x|+|\eta_y|)/2$.

In the limit $T\rightarrow 0$ we can treat $t_n=nT\rightarrow t$ as a continuous time variable and $\tilde H_{\rm int}^{\rm eff}(t)$ then determines the exact dynamics of the cavity QED system. Even for finite $T$, Eq.~\eqref{eq:effHamil2} serves as a  guideline to identify pulses that reproduce this dynamics to a very good approximation.  When $\eta_x=\eta_y$, $\tilde H_{\rm int}^{\rm eff}(t)$ corresponds to a JC model with detuning $\Delta_{\rm eff}$ and coupling strength $g_{\rm eff}$. In contrast, when $\eta_x=-\eta_y$, the effective Hamiltonian corresponds to a detuned anti-JC model, $\tilde H_{\rm int}^{\rm eff} \sim  (\sigma_+ a^\dag + \sigma_- a)$, while the case $\eta_y=0, \eta_x\neq0$ translates into a Rabi-coupling $\tilde H_{\rm int}^{\rm eff}\sim \sigma_x (a+a^\dag)$. Therefore, the identification of appropriate sequences of $X$ and $Y$ pulses not only allows us to recover the original JC interaction, but also to engineer effective interactions that were not present in the undriven system.

\subsection{Robust dynamical decoupling}
Apart from engineering the targeted effective coupling, a given pulse sequence must also fulfill its original purpose and protect the coupling against low-frequency noise and, ideally, also against pulse errors~\cite{Gullion:1990,Ali:2013}. This requirement imposes the following additional conditions on the modulation functions $f_{k}(t)$. First of all, in order to cancel unwanted frequency shifts, we require 
\begin{equation}
\tilde H^{\rm eff}_{\rm noise} = \frac{1}{2T}\int_0^T ds f_z(s)\xi(s)\sigma_z=0,
\end{equation} 
which for quasi-static noise, $T\ll \tau_c$, is achieved when
\begin{equation}\label{eq:condition_fz}
\gamma_z=\frac{1}{T}\int_0^T ds f_z(s) =0.
\end{equation} 
Second, small deviations from complete $\pi$-rotations will contribute with an effective Hamiltonian of the form
\begin{equation}\label{eq:qubitterms}
\delta\tilde H^{\rm eff}_{\rm drive}= \frac{\upsilon_{x}}{2}\sigma_x +\frac{\upsilon_{y}}{2}\sigma_y,
\end{equation}
where 
\begin{equation}
\begin{split}
\upsilon_{\eta=x,y}&=\frac{1}{T}\int_0^T ds f_{\eta}(s) \delta \Omega_{\eta}(s)=\frac{1}{T} \sum_i f_{\eta}(t_i^\eta) \delta \theta_i^{\eta}.
\end{split}
\end{equation}
In the same way as quasi-static shifts are canceled out by sequences that fulfill $\gamma_z=0$, sequences that fulfill
\begin{equation}
\gamma_{x,y}=\frac{1}{T}\int_0^T ds f_{x,y}(s) =0
\end{equation} 
will be robust against systematic rotation errors, $\delta\theta_i^{\eta}=\delta\theta^{\eta}\neq 0$. Note, however, that incomplete $\pi$-pulses can also be used on purpose to engineer deterministic Hamiltonian terms as in  Eq.~\eqref{eq:qubitterms} with $\upsilon_{\eta=x,y}\neq 0$. This approach can be combined with the requirement $\gamma_{x,y}=0$, for example, by applying alternating rotations with $\delta\theta_{i+1}^\eta=-\delta\theta_i^\eta$.  

\subsection{Protected JC and anti-JC models}
Let us now apply this general framework to identify specific pulse sequences that can be used to implement dynamically protected cavity QED Hamiltonians. 
We do so first for the case of the JC model and the anti-JC model, which only differ by a sign in the condition $\eta_x=\pm \eta_y$.  Note that when, for example, only $X$ pulses are applied we obtain $f_x(t)=1$ and either $\eta_y=0$ for $m=0$ or $\eta_x=0$ for $m>0$. Consequently, both $X$ and $Y$ pulses are required to engineer (anti) JC models with $|\eta_x|=|\eta_y|\neq0$. This explains the failure of the naive spin-echo pulse sequence discussed above. A second basic observation is that the functions $f_{x}(t)$ and $f_{y}(t)$ can only be periodic for an even number of $X$ or $Y$ pulses during the interval $T$. Therefore, the minimal sequence leading to an (anti) JC model must contain at least four (two $X$ and two $Y$) pulses. When both $f_{x}(t)$ and $f_{y}(t)$ are periodic, $f_{z}(t)$ is periodic as well.

\setlength{\arrayrulewidth}{1pt}   
\begin{table}[h!] 
\centering
\renewcommand{\arraystretch}{1.4} 
\setlength{\tabcolsep}{0pt}       

\begin{tabular}{ ||p{1.5cm}||p{1cm}|p{1cm}|p{1cm}|p{1cm}|p{1cm}|p{1cm}|}
 \hline
\fullcell{white}{1.1cm}{$m$} & \fullcell{white}{0.8cm}{0} & \fullcell{white}{0.8cm}{1} & \fullcell{white}{0.8cm}{2}& \fullcell{white}{0.8cm}{3} & \fullcell{white}{0.8cm}{4} & \fullcell{white}{0.8cm}{5}  \\ [0.5ex] 
 \hline\hline
\fullcell{white}{1.2cm}{XXYY} &
\fullcell{blue}{0.8cm}{0.5} &
\fullcell{IndianRed1}{0.8cm}{0.45} &
\fullcell{blue}{0.8cm}{0.32} &
\fullcell{IndianRed1}{0.8cm}{0.15} & 
\fullcell{gray}{0.8cm}{0} &
\fullcell{IndianRed1}{0.8cm}{0.09} \\
\hline
\fullcell{white}{1.2cm}{XY8} &
\fullcell{gray}{0.8cm}{0} &
\fullcell{IndianRed1}{0.8cm}{0.34} &
\fullcell{blue}{0.8cm}{0.45} &
\fullcell{IndianRed1}{0.8cm}{0.28} & 
\fullcell{gray}{0.8cm}{0} &
\fullcell{IndianRed1}{0.8cm}{0.17} \\
\hline
\fullcell{white}{1.2cm}{XX} &
\fullcell{PaleGreen1}{0.8cm}{0.5} &
\fullcell{yellow}{0.8cm}{0.32} &
\fullcell{gray}{0.8cm}{0} &
\fullcell{yellow}{0.8cm}{0.11} & 
\fullcell{gray}{0.8cm}{0} &
\fullcell{yellow}{0.8cm}{0.06} \\
\hline
\fullcell{white}{1.2cm}{YY} &
\fullcell{yellow}{0.8cm}{0.5} &
\fullcell{PaleGreen1}{0.8cm}{0.32} &
\fullcell{gray}{0.8cm}{0} &
\fullcell{PaleGreen1}{0.8cm}{0.11} & 
\fullcell{gray}{0.8cm}{0} &
\fullcell{PaleGreen1}{0.8cm}{0.06} \\
\hline
\end{tabular}
\caption{\ia{Summary of the numerical values of $\eta$, which determines the effective coupling strength, $g_{\rm eff}=g\eta$, as obtained for different values of the detuning $\Delta =2\pi m/T$ and different pulse sequences. The blue and red colors indicate that the corresponding pulse sequences realize JC and anti-JC interactions, respectively. The green and yellow colors refer to Rabi interactions of $\sigma_x$-type and $\sigma_y$-type instead.}
}\label{tab:couplings}
\end{table}

\subsubsection{Four-pulse sequence}

In the most general four-pulse sequence $P^{(1)}_{t_1}P^{(2)}_{t_2}P^{(3)}_{t_3}P^{(4)}_{t_4}$, we can still choose the times $t_i$ and the order of the pulses $P=X,Y$ to apply. For simplicity, here we focus on sequences of equally spaced pulses with a spacing $\tau=T/4$. Since every pulse flips the sign of $f_z(t)$, this choice ensures that the condition $\gamma_z=0$ is satisfied and the sequence protects against low-frequency noise. The condition $\eta_y=\pm \eta_x$, required for the (anti) JC model, then translates into the relation $f_y(t)=f_x(t+T/2)$ for the toggling functions. From these assumptions it follows that $\eta_{y}=(-1)^m\eta_x$, where $m\in \mathbb{Z}$ is determined by the choice of the detuning $\Delta=2\pi m/T$.

The only four-pulse sequence that satisfies all those constraints is $X_{t_1}X_{t_1+\tau}Y_{t_1+2\tau}Y_{t_1+3\tau}$, where the time-offset $t_1\in [-\tau,\tau]$ is a free parameter. Note that here we allow for negative values of $t_1$ to represent the sequence $X_{t_1+\tau}Y_{t_1+2\tau}Y_{t_1+3\tau}X_{t_1+4\tau}$. For this set of pulses  we obtain the effective parameters $\eta_y=\eta e^{i\phi_y}$ and $\eta_x=(-1)^m\eta_y$, where
\begin{equation}
\eta= \frac{2}{m \pi} \sin\left(\frac{m \pi}{4}\right)
\end{equation}
and $\phi_y=\pi m (8 t_1/T+1)/4-\pi$ for $m\neq0$, and $\eta=0.5$ and $\phi_y=0$ for $m=0$. The relative phase between $\eta_x$ and $\eta_y$ can be either $0$ or $\pi$, such that the effective interaction has the form of a JC model,
\begin{equation}\label{eq:effectiveJC}
\tilde H_{\rm int}^{\rm eff} = \eta g \left(\sigma_-  a^\dagger e^{i\phi}  +{\rm H.c.} \right),
\end{equation}
with $\phi=\phi_y$ when $m$ is even and that of an anti-JC model,
\begin{equation}
\tilde H_{\rm int}^{\rm eff} = \eta g \left(\sigma_+ a^\dagger e^{i\phi}  +{\rm H.c.} \right),
\end{equation}
with $\phi=\phi_y+\pi$ when $m$ is odd. Both models can be generalized to include a finite detuning $\Delta_{\rm eff}$, as long as $|\Delta_{\rm eff}|  \ll 2\pi  /T$.

In Table~\ref{tab:couplings} we summarize the values of the coupling coefficient $\eta$ obtained for different $m\in \mathbb{Z}$, as well as the type of the resulting light-matter interaction. Interestingly, for values of $m$ that are multiples of $4$, the interaction cancels out completely. Such pulse sequences can be used, for example, to decouple the TLS from the cavity mode during idle times. In general, we find that larger values of $|m|$ result in a smaller effective coupling strength. However, depending on the applied pulse sequence, a minimal value $m\neq0$ might be required to obtain a specific effective model. In this case, the initial detuning between the TLS and the cavity mode,
\begin{equation}
\Delta \simeq \frac{2\pi m}{T},
\end{equation}
must match the pulse period. Importantly, this relation implies that by adjusting $\Delta$ accordingly, the pulse period $T$ can be set to be arbitrarily short without reducing the coupling strength.

\ia{Prior to this work, Groszkowski {\it et al.}~\cite{Groszkowski:2022} suggested another four-pulse sequence to protect a resonant JC-coupling. In our notation, this sequence can be written as $X_{T/2} X_{3T/4}Y_{3T/4} Y_T$. It is not equally spaced and at time $3T/4$ a combination of an $X$ and a $Y$ pulse is applied right after each other to implement an effective $Z$-rotation. This sequence satisfies all the conditions from above and realizes an effective JC-coupling with $g_{\rm eff}=g/2$. For an ideal and noiseless systems, this sequence is very robust with respect to a finite spacing between the pulses, but typically it performs worse than the other sequences presented here under realistic conditions. This behavior can be understood from the error analysis presented in Appendix~\ref{app:ErrorEstimates}.}

\subsubsection{Eight-pulse sequence}
A practical drawback of the four-pulse sequence from above is that it does not satisfy the conditions $\gamma_{x,y}=0$, which are required to make it robust against pulse imperfections. Therefore, we now consider the XY8 sequence~\cite{Gullion:1990}  $X_{t_1}Y_{t_1+\tau}X_{t_1+2\tau}Y_{t_1+3\tau}Y_{t_1+4\tau}X_{t_1+5\tau}Y_{t_1+6\tau}X_{t_1+7\tau}$ with a pulse spacing $\tau=T/8$ and $-\tau<t_1<\tau$. Again, we find that this sequence fulfills the  symmetry condition $f_y(t)=f_x(t+T/2)$, which results in an effective (anti) JC interaction. The coefficients in this case are $\eta_y=\eta e^{i\phi_y}$ and $\eta_x=(-1)^m\eta_y$ with
\begin{equation}
\eta =\frac{4}{\pi m} \sin\left(\frac{\pi m}{4}\right)\cos\left(\frac{5\pi m}{8}\right),
\end{equation}
and $\phi_y=\pi m (16t_1/T-1)/8 +\pi(m+1)$, while $\eta=0$ for $m=0$. As in the previous case, even values of $m$ will generate a JC interaction, odd values will generate an anti-JC interaction. For values of $m$ that are multiples of four, the light-matter interaction cancels out.

From Table~\ref{tab:couplings} we see that the XY8 sequence achieves similar values for the effective coupling strength as the XXYY sequence. At the same time, however, it satisfies $\gamma_{x}=\gamma_y=0$ and it is thus much more robust with respect to pulse imperfections.

\begin{figure*}
	\includegraphics[width=2\columnwidth]{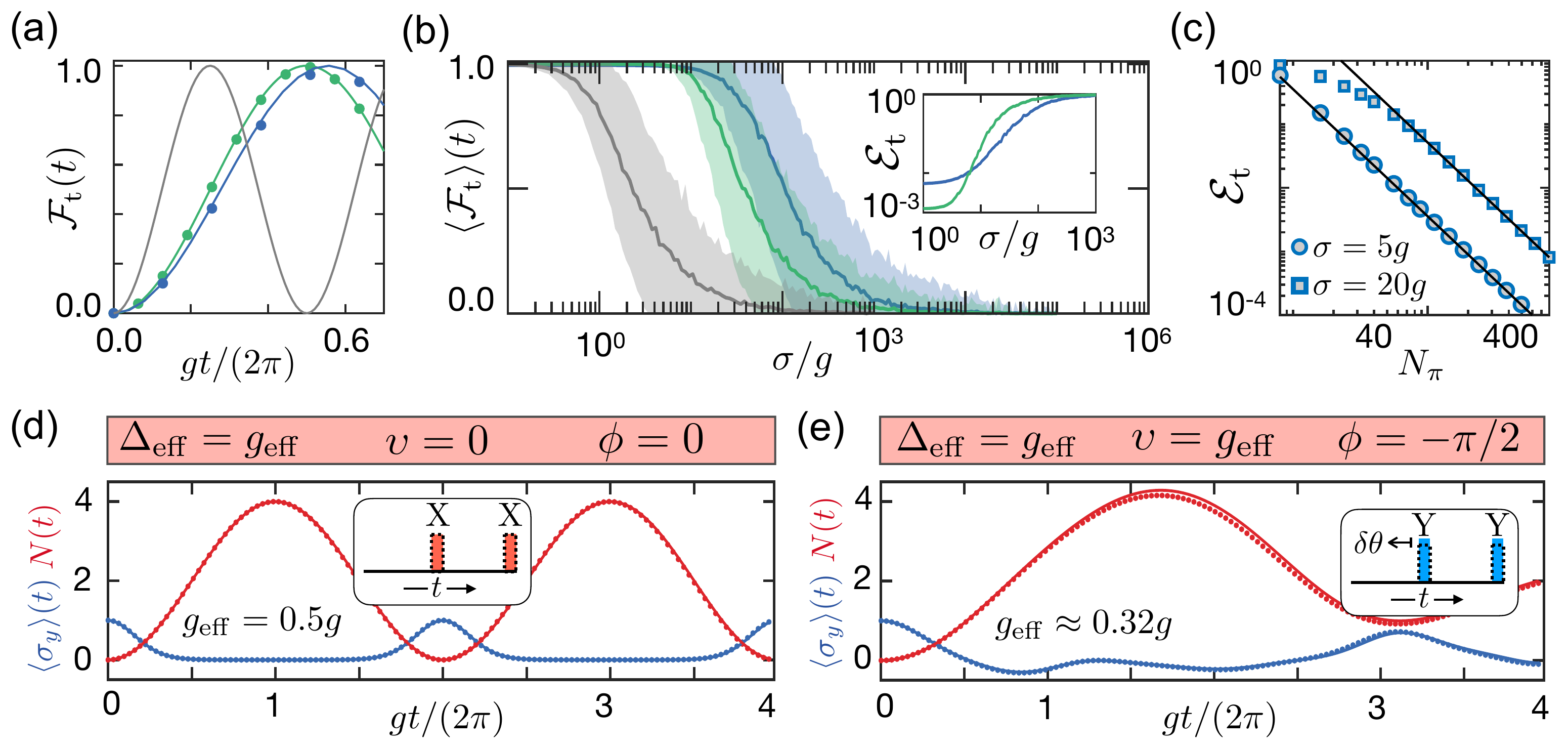}
	\caption{ (a) Plot of the transfer fidelity as a function of time for the pulse sequences XY8$_{m=2}$  (blue) and XXYY$_{m=0}$ (green) in the absence of noise. \ia{The markers indicate the result of exact numerical simulations with an} interpulse spacing $\tau=0.1g^{-1}$, a pulse width $\tau_\pi=0.1\tau$ and $t_1=\tau/2$, \ia{while solid lines follow an ideal evolution with the corresponding $g_{\rm eff}$}. The shaded line represents the transfer fidelity for the bare JC-interaction.  (b) For the same sequences the average transfer fidelity at time $T_{\rm t}=\pi/(2g_{\rm eff})$ is plotted as a function of the noise strength $\sigma$ and for a total of $N_\pi=64$ pulses. 
    While the solid lines represent the average values obtained as a result of 1000 independent noise realizations, randomly sampled from the probability distribution $P(\xi)=(2\pi\sigma^2)^{-1/2}\exp{(-\xi^2/2\sigma^2)}$, the shaded area indicates its variation (one standard deviation). The inset shows the corresponding transfer error $\mathcal{E}_{\rm t}$ on a logarithmic scale. (c) Dependence of the average transfer error on the  number of pulses $N_\pi$ for the XY8$_{m=2}$ sequence with $\sigma=5g$ (round markers) and $\sigma=20g$ (square markers). The solid lines indicate the analytic prediction from Eq.~\eqref{eq:XY8Error} for the same parameters. (d) Time evolution of the observables $\langle\sigma_y\rangle$ and $N=\langle a^\dagger a\rangle$ for the XX$_{m=0}$ sequence with parameters $\tau=0.1g^{-1}$, $t_1=\tau$, $\tau_\pi=10^{-2}\tau$, and $\Delta=g_{\rm eff}$. The markers indicate the result obtained from exact numerical simulations, while the solid lines follow the prediction of the effective model in Eq.~\eqref{eq:QRM} with $\Delta_{\rm eff}=g_{\rm eff}$, $\upsilon=0$, and $\phi=0$. (e) The same as in (d) but for the sequence YY$_{m=1}$ with a cavity detuning, $\Delta=2\pi /T+g_{\rm eff}$ and pulse area $\pi+\delta \theta$, where $\delta\theta=g_{\rm eff} T/2$. This simulation is compared to the effective model in Eq.~\eqref{eq:QRM} with $\Delta_{\rm eff}=\upsilon=g_{\rm eff}$ and $\phi=-\pi/2$. \ia{The small mismatch with the effective model, most visible at the peak values of $N(t)$, is reduced when using a smaller interpulse spacing $\tau$.} The insets in panels (d) and (e) illustrate the pulse sequence, where the area enclosed by the dotted lines corresponds to $\pi$. Notably, in panel (e), the pulse area exceeds $\pi$ by a small amount $\delta\theta$. }
\label{Fig:3}
\end{figure*}

\subsection{Quantum Rabi model}\label{subsec:QRM}

As already pointed out above, some pulse sequences give rise to effective couplings of the form $\tilde{H}_{\rm int}^{\rm eff} \sim \sigma_x (a + a^\dag)$. This interaction plays a prominent role for various quantum control schemes~\cite{Sorensen:2000,Qin:2024}, but it also appears in the modeling of light-matter interactions in the so-called ultrastrong-coupling regime~\cite{Forn:2019,Frisk:2019}, where the usual rotating-wave approximation is no longer applicable.

The simplest sequences giving rise to such a coupling are $X_{t_1}X_{t_2}$ for $m=0$ and $Y_{t_1}Y_{t_2}$ for $m\neq0$. In both cases, the condition $\gamma_z=0$ requires that  $t_2=t_1+T/2$, and  $t_1$ is the only free parameter. For $X_{t_1}X_{t_1+T/2}$ and $m=0$,
the effective parameters are $\eta_y=0$ and $\eta_x=1$. The effective interaction then assumes the anticipated form, 
\begin{equation}
\tilde H_{\rm int}^{\rm eff} = \eta g \left(  a^\dagger e^{i\phi }  +{\rm H.c.} \right)\sigma_x,
\end{equation} 
with $\eta=0.5$ and $\phi=0$. The same is true for the $Y_{t_1}Y_{t_1+T/2}$ sequence and $m$ odd, but with parameters
\begin{eqnarray}
\eta=\frac{1}{\pi m}\sin^2(\pi m/2) 
\end{eqnarray}
and $\phi=\pi m( 2t_1/T+1)+\pi/2$. Table~\ref{tab:couplings}(c) summarizes the values of $\eta$ obtained for different $m$ for the $XX$ and the $YY$ sequence. 

To realize the full quantum Rabi model, we can choose a finite effective detuning $\Delta_{\rm eff}$ and increase the rotation angle of each pulse to $\pi+\delta \theta$ with $\delta \theta \ll \pi$. In the case of the $Y_{t_1}Y_{t_1+T/2}$ sequence with $m$ odd, the resulting effective Hamiltonian is then given by  
\begin{equation}\label{eq:QRM}
\tilde H_{\rm int}^{\rm eff} = \frac{\upsilon}{2}\sigma_y+ \eta g \left(  a^\dagger e^{i\phi } e^{i\Delta_{\rm eff}t}  +{\rm H.c.} \right)\sigma_x,
\end{equation} 
where $\upsilon=2\delta\theta/T$. In the respective rotating frame, Eq.~\eqref{eq:QRM} describes the coupling of an effective TLS with frequency $\omega_0\equiv \upsilon$ to a cavity mode of frequency $\omega_c\equiv \Delta_{\rm eff}$ and coupling strength $g_{\rm eff}=\eta g$. These parameters can be tuned independently through an appropriate choice of pulse parameters, enabling the system to reach the ultrastrong coupling regime, $g_{\rm eff} \gtrsim \omega_0, \omega_c$, within this effective description.

\subsection{Examples and Performance}\label{subsec:ExamplesPerformance}
To illustrate the effectiveness of these DD strategies, Fig.~\ref{Fig:3} shows the results of numerical simulations of the dynamics of the full cavity QED system for a finite number of echo pulses and in the presence of static frequency shifts of varying strength $\sigma$. In Fig.~\ref{Fig:3}(a) we first compare the evolution of the original JC model with that of the effective JC model obtained using the XXYY$_{m=0}$ and the XY8$_{m=2}$ sequences in the absence of noise. In all three cases we assume a resonant (effective) coupling and we plot the transfer fidelity $\mathcal{F}_{\rm t}(t)$ as a figure of merit. We see that already for a pulse spacing of $\tau=0.1g^{-1}$, the effective evolution agrees very well with the analytic predictions from above and the effective coupling parameters listed in Table~\ref{tab:couplings}. 

Including the influence of quasi-static noise, Fig.~\ref{Fig:3}(b) demonstrates the increased robustness obtained for both pulse sequences for a total of $N_\pi=64$ $\pi$-pulses. Under the same conditions, the inset shows the residual transfer error, $\mathcal{E}_{\rm t}=1-\langle\mathcal{F}_{\rm t}(t=T_{\rm t})\rangle$, where $T_{\rm t}=\pi/(2g_{\rm eff})$ now refers to the adjusted transfer time. We see that the error is reduced to values below $10^{-2}$, even when the noise strength exceeds the bare coupling strength. However, for small noise we also observe a saturation of the error, which arises from the finite pulse interval $\tau$ and depends on the pulse sequence. As we discuss in more detail in Appendix~\ref{App:FullEff} and~\ref{app:ErrorEstimates}, we can use second-order perturbation theory to account for those effects. Focusing on the XY8$_{m=2}$ sequence, which is more robust with respect to pulse errors and a finite width of the pulses, we obtain the transfer error
\begin{equation}\label{eq:XY8Error}
\mathcal{E}_{\rm t} \approx \left[\Big(\frac{2\sigma}{3\pi}\Big)^2+ 1.88 g^2\right] \tau^2 ,
\end{equation} 
where  $\tau=T_{\rm t}/N_\pi$. This result shows that the pulse spacing $\tau$ must be short compared to both the inverse noise strength and the inverse bare coupling constant. Once this condition is achieved, the error decreases very rapidly with the number of pulses, $\mathcal{E}_{\rm t}\propto N_\pi^{-2}$. When $\sigma \gg g$, the second term in Eq.~\eqref{eq:XY8Error} becomes negligible and the suppression of the error compared to the evolution without DD pulses [see Eq.~\eqref{eq:bare_fidelity}] is approximately given by
\begin{equation}\label{eq:ErrorRatio}
\frac{\mathcal{E}_{\rm t}|_{\rm DD}}{\mathcal{E}_{\rm t}|_{\rm no DD}} \approx \left(\frac{2}{3\eta N_\pi}\right)^{2}.
\end{equation} 
This scaling is very accurately reproduced by the exact numerical simulations shown in Fig.~\ref{Fig:3}(c).

Finally, in Fig.~\ref{Fig:3}(d) and Fig.~\ref{Fig:3}(e) we also illustrate the implementation of the Rabi model under two slightly different conditions. In the first plot we consider an XX$_{m=0}$ sequence with a finite $\Delta_{\rm eff}=g_{\rm eff}$. In the second plot, we consider a YY$_{m=1}$ sequence, also with $\Delta_{\rm eff}=g_{\rm eff}$, but with an additional rotation angle $\delta \theta= g_{\rm eff} T/2 >0$. As explained in Sec.~\ref{subsec:QRM}, the latter corresponds to an effective level splitting of $\upsilon=g_{\rm eff}$ for the TLS. In both cases we find an excellent agreement between the exact dynamics and the one predicted by the effective model. Note, however, that for the simulation of the quantum Rabi model we have assumed a pulse length of $\tau_\pi=10^{-2}\tau$, which is an order of magnitude shorter than for the simulations of the JC model. This is due to the fact that the XX and YY sequences are more sensitive to pulse-width effects than the XXYY and XY8 sequences.

\section{Protecting cavity-mediated spin-spin interactions}

In many cavity QED experiments, the primary purpose of the cavity mode is to mediate coherent interactions between otherwise decoupled TLSs. To study such applications, we extend our model to a scenario with two TLSs that are coupled to the same cavity mode with coupling strengths $g_j$, where $j=1,2$. Assuming identical bare transition frequencies,
the resulting Hamiltonian reads
\begin{equation}\label{eq:TCHamil}
H(t)=\Delta a^\dagger a+ \sum_{j=1}^{2}\frac{\xi_j(t)}{2}\sigma_j^z +\sum_{j=1}^{2}g_j (\sigma_j^+a+\sigma_j^-a^\dagger ),
\end{equation}
where the $\xi_j$ are independent noise processes. When $|\Delta| \gg g_j,\xi_j$, the coupling to the cavity mode can be treated in perturbation theory and we obtain the following effective Hamiltonian
\begin{equation}\label{eq:eff2QHamil}
H(t)\approx \sum_{j=1}^{2}\frac{\hat \xi_j(t)}{2}\sigma_j^z - J(\sigma_1^+\sigma_2^- +\sigma_1^-\sigma_2^+),
\end{equation}
where $\hat \xi_j(t) = \xi_j(t)-\frac{|g_j|^2}{\Delta} (2a^\dagger a+1)$ and $J=g_1g_2/\Delta$.  In this detuned limit, the second term in Eq.~\eqref{eq:eff2QHamil} exchanges excitations between the TLSs, while only virtually populating the cavity mode. Specifically, after a time $T_{\rm e}=\pi/(4J)$, this term transforms the product state $|eg\rangle$ into the maximally entangled state $|\Psi\rangle=\frac{1}{\sqrt{2}}(|eg\rangle+i|ge\rangle)$, which is relevant for many quantum information processing applications. In the following we focus on this specific entanglement operation and use the entanglement fidelity
\begin{equation}
\mathcal{F}_{\rm e} = {\rm Tr}\{ |\Psi\rangle\langle\Psi| \rho(T_{\rm e})\},
\end{equation} 
where $\rho(t)$ is the full density operator of the cavity QED system, to quantify the accuracy of cavity-mediated interactions.

Similar to the quantum state transfer analyzed in the previous section, the quality of the achieved spin-spin entanglement will decrease in the presence of uncontrolled fluctuations, $\xi_j(t)$. In Fig.~\ref{Fig:4}(a) the grey line shows the average fidelity $\langle\mathcal{F}_{\rm e}\rangle$ as a function of the noise strength $\sigma$. For concreteness, we have assumed $g_1=g_2=g$, $\Delta=30 g$, a cavity in the ground state, and independent, static fluctuations with $\langle \xi_i\xi_j\rangle=\sigma^2 \delta_{ij}$.
In this static limit, the noise-averaged fidelity decays as 
\begin{equation}
 \langle\mathcal{F}_{\rm e}\rangle\approx 1-\frac{1}{8}\left(\frac{\sigma}{J}\right)^2,
\end{equation}
and, therefore, the noise must be weak compared to the effective coupling $J$ in order generate significant entanglement. 
Note that for a thermally populated cavity mode, the fluctuating AC Stark shift $\sim a^\dag a $ induces an additional random frequency shift with a strength $\sigma_{\rm th} = g^2 n_{\rm th} /|\Delta| $, where $n_{\rm th}$ is the thermal equilibrium occupation number of the mode~\cite{Arrazola:2024}. However, to simplify the following discussion, we assume $n_{\rm th}=0$ and do not take this term explicitly into account.

\subsection{Large-detuning regime}

In contrast to the resonant JC coupling discussed in Sec.~\ref{subsec:vaccumRabi}, the effective Hamiltonian in Eq.~\eqref{eq:eff2QHamil} is already consistent with  DD techniques. Specifically, the application of fast $\pi$-rotations on both TLSs simultaneously does not take the combined state out of the subspace of interest, $\{|eg\rangle,|ge\rangle\}$, and is compatible with the discussed entanglement generation protocol. This can be understood by rewriting Hamiltonian~\eqref{eq:eff2QHamil} in the toggling frame. According to Eq.~\eqref{eq:fliprules}, in this frame the flip-flop term remains invariant, 
$\tilde{\sigma}_1^+(t)\tilde{\sigma}_2^-(t) +{\rm H.c.}=\sigma_1^+\sigma_2^- +{\rm H.c.}$, as long as the $\pi$-rotations are applied to both TLSs simultaneously. The resulting Hamiltonian is then given by
\begin{equation}\label{eq:togglingflipflop}
\tilde{H}(t)\approx \sum_{j=1}^{2}\frac{\hat \xi_j(t)}{2}f_z(t)\sigma_j^z - J(\sigma_1^+\sigma_2^- +\sigma_1^-\sigma_2^+),
\end{equation}
and for any pulse sequence with $\gamma_z=0$, the static noise can be canceled to first order without affecting the interaction term.

\begin{figure}
	\includegraphics[width=\columnwidth]{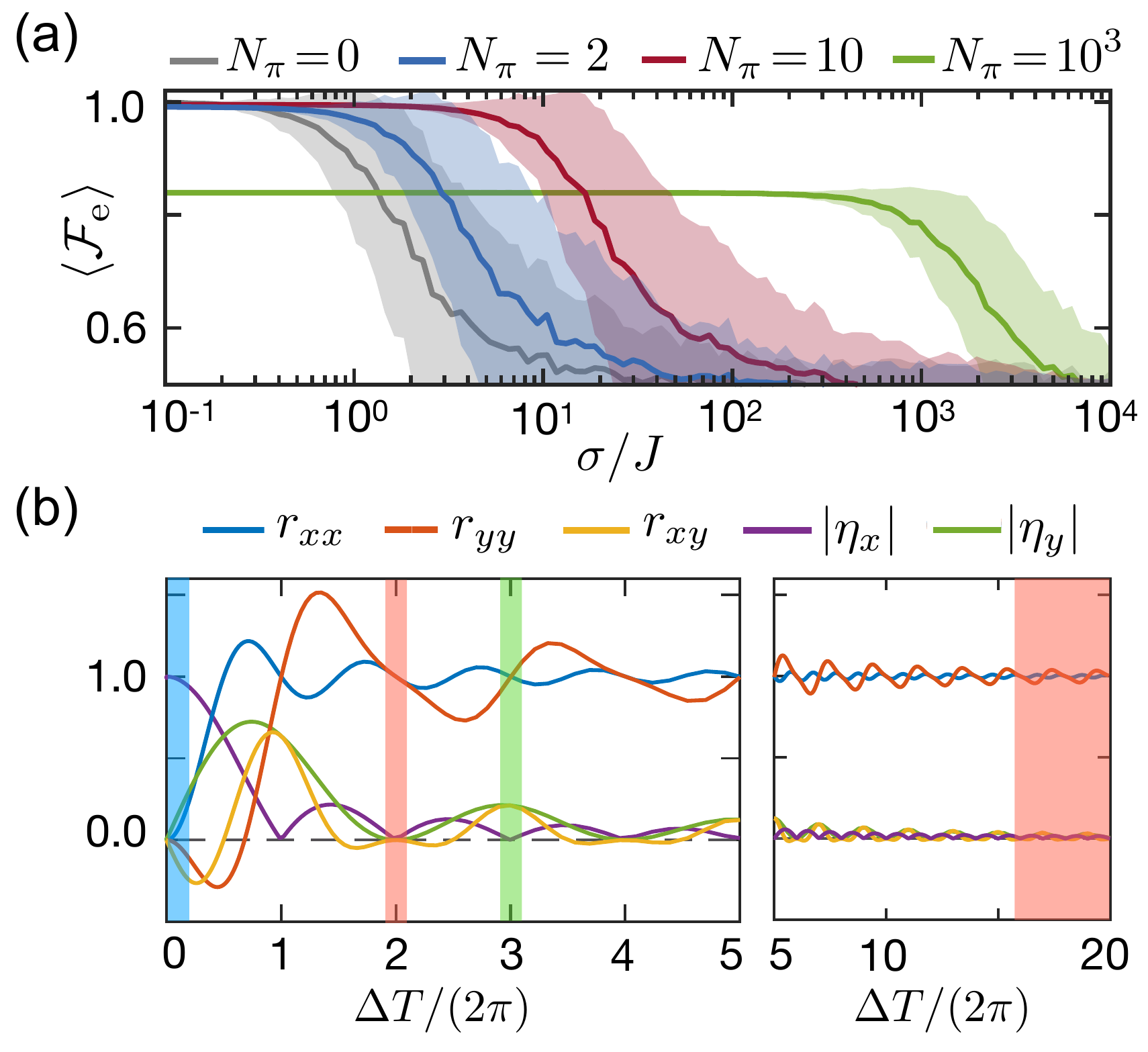}
	\caption{(a) Plot of the average entanglement fidelity $\langle \mathcal{F}_{\rm e}\rangle$ as a function of the noise strength $\sigma$ for a cavity QED system with two TLSs. The TLSs are detuned by $\Delta= 30g$ and undergo cavity-mediated flip-flop interactions with strength $J=g^2/\Delta$. The solid lines (shaded areas) represent the average values (standard deviations) obtained  from an average over 500 noise realizations for  different numbers of $N_\pi$ instantaneous $\pi$-pulses. (b) Dependence of the effective interaction parameters on $\Delta$ for the $X_{T/2}X_T$ pulse sequence. The red, blue and green shaded stripes indicate the regimes leading to flip-flop, Ising and squeezing interactions, respectively. See Sec.~\ref{subsec:EffectiveSS} and Fig.~\ref{Fig:5} for more details.} 
	\label{Fig:4}
\end{figure}

In Fig.~\ref{Fig:4}(a), the blue line shows the resulting average fidelity $\langle\mathcal{F}_{\rm e}\rangle$ after introducing only a single $\pi$-rotation (on both TLSs) in the middle and at the end of the evolution. As expected, we see that the performance of the echoed case is significantly better than without the decoupling pulses. Note, however, that operation is still affected by noise, despite considering purely static fluctuations, $\xi_j(t)=\xi_j$, and ideal $\pi$-pulses. This is in stark contrast to more commonly investigated spin systems with Ising interactions of the form $J\sigma_1^z\sigma_2^z$~\cite{Vandersypen:2005}, where a single $\pi$-pulse applied to each TLS would be sufficient to cancel any static frequency shifts exactly. The difference in the current cavity setup arises from the fact that the noise and the interaction term in Eq.~\eqref{eq:togglingflipflop} do not commute and thus the overall dynamics is more involved.  

This observation suggests that in order to improve cavity-mediated interactions, the interpulse spacing must satisfy not only $\tau \lesssim \tau_c$, but also $\tau\lesssim 1/\sigma$. Indeed, in Fig.~\ref{Fig:4}(a), the red line shows the fidelity for an evolution interrupted by a total of ten $X$-rotations of each TLS. In this case, the interpulse spacing $\tau$ is ten times shorter than in the previous example (blue line), increasing the level of tolerable noise by an order of magnitude.

However, the behavior discussed so far is valid only when the period $T$ is long compared to $\Delta^{-1}$, i.e. $\Delta T \gg1$. In this case, it is sufficient to apply the toggling-frame transformation to Hamiltonian~\eqref{eq:eff2QHamil} instead of the original JC Hamiltonian in Eq.~\eqref{eq:TCHamil}. For example, in the sequences with up to ten pulses discussed above, we have $\Delta T/(2\pi)\gtrsim 22$. In contrast, the green line in Fig.~\ref{Fig:4}(a) shows a sequence with a total of 1000 pulses applied to each TLS, in which case $\Delta T/(2\pi)\approx 0.22$. In this regime, the effect of the noise is suppressed up to very high levels, but already for very low noise amplitudes, the fidelity $\mathcal{F}_{\rm e}$ is significantly degraded. We conclude that while the application of more and more pulses makes the evolution more robust, it also changes the form of the effective interaction, and the system can no longer be modeled by the simple flip-flop interaction as given in Eq.~\eqref{eq:togglingflipflop}.

\subsection{Arbitrary detunings}\label{sec:ArbitraryDetuning}
To model cavity-mediated interactions in a regime where the period $T$ is comparable or shorter to $\Delta^{-1}$,
the transformation to the toggling frame must be applied to the original Hamiltonian in Eq.~\eqref{eq:TCHamil}, as it was done in Eq.~\eqref{eq:TogglinFrameHamil} for a single TLS. Only after this transformation, we can use a second-order Magnus expansion to derive the time-averaged Hamiltonian (see Appendix~\ref{App:FullEff}), 
\begin{equation}\label{eq:effHamil0}
\tilde H_{\rm eff}(nT)=\tilde H^{(1)}_{\rm eff}(nT)+\tilde H^{(2)}_{\rm eff}(nT),
\end{equation}
which describes the effective evolution of the whole system during the $n$-th time interval $\{n T,(n+1)T\}$. Here, the first-order and second-order effective interactions are given by 
\begin{equation}\label{eq:1stExp}
\tilde H^{(1)}_{\rm eff} (nT) = \frac{1}{T} \int_{n T}^{(n+1) T} ds \tilde H(s)
\end{equation} 
and 
\begin{equation}\label{eq:2ndExp}
\tilde H^{(2)}_{\rm eff}(nT) = \frac{-i}{2T} \int_{n T}^{(n+1)T} dt \int_{nT}^tds \, [\tilde H(t), \tilde H(s)],
\end{equation}
respectively~\cite{Brinkmann:2016}. For convenience, we reorganize Eq.~\eqref{eq:effHamil0} as
\begin{equation}\label{eq:effHamil1}
\tilde H_{\rm eff}(nT)=\tilde H^{(1)}_{\rm sc}(nT)+\tilde H^{(2)}_{\rm ss}+\tilde H_{\rm corr}(nT) ,
\end{equation}
where the first two terms represent the dominating effective interactions, while the third term contains additional unwanted contributions from the noise and other imperfections. These corrections will be discussed in more detail in Sec.~\ref{subsec:ErrorModel} below and in Appendix~\ref{app:ErrorEstimates}.

With the same conventions as introduced in Sec.~\ref{sec:ProtectedJC} for a single TLS, we obtain a spin-cavity coupling 
\begin{equation}\label{eq:1stHamil}
\tilde H^{(1)}_{\rm sc}(t) = \sum_{j=1}^2\frac{g_j}{2}\left[(\eta_x  \sigma_j^x - i \eta_y \sigma_j^y)   a^\dagger e^{i\Delta_{\rm eff} t} +{\rm H.c.} \right]
\end{equation} 
and a second-order spin-spin interaction of the form  \begin{equation}\label{eq:HamilManyPulse}
\tilde H^{(2)}_{\rm ss} = - \frac{J}{2} \sum_{\{u,v\}=\{x,y\}} r_{uv}\, \sigma_1^u \sigma_2^v. 
\end{equation}
Here, $J=g_1g_2/\Delta$ and we have introduced the additional dimensionless coefficients
\begin{eqnarray}
r_{uv}=-\frac{\Delta}{T}{\rm Im}\left[\int_0^T dt\int_0^t  ds  \, \tilde{f}^*_u(t)  \tilde f_v(s)e^{-i\Delta (t-s)}\right],
\end{eqnarray}
where $\tilde{f}_{x}(t)={f}_{x}(t)$, $\tilde{f}_{y}(t)=-i{f}_{y}(t)$. For illustration, we consider in Fig.~\ref{Fig:4}(c) the simple pulse sequence $X_{T/2}X_{T}$ and plot the values of $\eta_{x,y}$ and the coefficients $r_{xx}$, $r_{yy}$ and $r_{xy} = r_{yx}$ as a function of the detuning $\Delta$.

\subsection{Effective spin-spin interactions}\label{subsec:EffectiveSS}
From the form of the two contributions of $\tilde H_{\rm eff}$ given in Eq.~\eqref{eq:1stHamil} and Eq.~\eqref{eq:HamilManyPulse}, we see that there are different ways to obtain effective interactions between the TLSs. First, one can engineer an appropriate first-order Hamiltonian $\tilde H^{(1)}_{\rm sc}$, with an effective detuning $\Delta_{\rm eff}$ that is large compared to the effective coupling $g_{\rm eff}\sim \eta g$. Similar to an unperturbed cavity QED system, this will generate cavity-mediated interactions with a scaling $\sim g_{\rm eff}^2/\Delta_{\rm eff}$. In addition, one can directly make use of the second-order Hamiltonian $\tilde H^{(2)}_{\rm ss}$, which represents an additional independent contribution that scales as $g^2/\Delta$. This combination, together with the strong dependence of all the coefficients on the detuning and the chosen pulse sequence, offers a large flexibility for engineering cavity-mediated interactions that are at the same time protected against noise. In the following, we illustrate these possibilities in terms of a few basic examples for the $X_{T/2}X_{T}$ sequence.

\subsubsection{Flip-flop interactions}
We start with the implementation of the flip-flop interaction
\begin{equation}\label{eq:flipflopHamil}
\tilde H_{\rm ss}^{(2)}\simeq - J (\sigma_1^+\sigma_2^-+ \sigma_1^-\sigma_2^+).   
\end{equation}
To obtain this form, we require $r_{xx,yy}\rightarrow 1$ and $r_{xy}, |\eta_{x,y}|\rightarrow 0$. As we can see from Fig.~\ref{Fig:4}(c), this is always satisfied in the large detuning limit, $\Delta T \gg 1$. However, we can also identify specific conditions, $\Delta = m\times2\pi/T$ with $m=2,4,...$, where $r_{xx,yy}= 1$ while $r_{xy}$ and $\eta_{x,y}$ vanish. Thus, in these cases, the effective flip-flop interaction can be obtained already for small and moderate detunings or, equivalently, for short interpulse spacings. Note that this feature is not unique to the $X_{T/2}X_{T}$ sequence and similar conditions can also found for more complicated pulse sequences.

\subsubsection{Ising interactions}
Another relevant type of coupling is the Ising interaction $\sim \sigma_1^x\sigma_2^x$. As we can see from Fig.~\ref{Fig:4}(c), there is no value for the detuning $\Delta$ for which only $r_{xx}$ is non-vanishing. Therefore, in this case, we apply a different approach and set $\Delta=\Delta_{\rm eff}$, with $\Delta_{\rm eff}T\ll 1$. In this limit the first-order coupling dominates and the effective Hamiltonian is 
\begin{equation}\label{eq:MSHamil}
\tilde H_{\rm sc}^{(1)}(t)\simeq g_{\rm eff} (a^\dag e^{i\phi_x} e^{i\Delta_{\rm eff}t} + {\rm H.c.}) S_x
\end{equation}
with $g_{\rm eff}=|\eta_x|g/2$ and $S_x=\sum_j (g_j/g) \sigma_j^x$.
If, in addition to $\Delta_{\rm eff} T\ll 1$, we choose $g_{\rm eff}\ll \Delta_{\rm eff}$, the effect of  $\tilde H_{\rm sc}^{(1)}$ is well described by the spin-spin Hamiltonian 
\begin{equation}
\tilde H_{\rm ss}^{(1)}\simeq -\frac{g_{\rm eff}^2}{\Delta_{\rm eff}}S_x^2,   
\end{equation}
which can be derived from Eq.~\eqref{eq:MSHamil} using second-order perturbation theory.

\subsubsection{Squeezing interactions}
Finally, another relevant application is the implementation of squeezing interactions $\sim (\sigma_1^+\sigma_2^++ \sigma_1^-\sigma_2^-)$. To do so we combine the two strategies from above and fix the detuning as $\Delta=2\pi m/T +\Delta_{\rm eff}$, with $m=3$ and $\Delta_{\rm eff }\ll\Delta$. In this regime, the second-order Hamiltonian $\tilde H^{(2)}_{\rm ss}$ follows a flip-flop interaction, while the first-order term $\tilde H^{(1)}_{\rm sc}$ contributes the effective Ising interaction $\tilde H_{\rm ss}^{(1)}\simeq -g_{\rm eff}^2/\Delta_{\rm eff} S_y^2 $ with $g_{\rm eff}=|\eta_y|g/2$. If $\Delta_{\rm eff}$ is chosen to be $\Delta_{\rm eff}=-\pi m|\eta_y|^2/T$, the combination of these two distinct interactions results in 
\begin{equation}
\tilde H_{\rm ss}\simeq \tilde{J}\sigma_1^+  \sigma_2^+ + \tilde{J}^*\sigma_1^-\sigma_2^-
\end{equation}
with $\tilde J=-J(1-i|\eta_y|)$. When generalized to multiple TLSs, this Hamiltonian is equivalent to a two-axis squeezing Hamiltonian $S_x^2-S_y^2$, as relevant for quantum sensing applications~\cite{Degen:2017}. \ia{This squeezing interaction will directly benefit from the dynamical protections, as we will discuss in Sec.~\ref{subsec:Coop} below.}

\begin{figure}
	\includegraphics[width=\columnwidth]{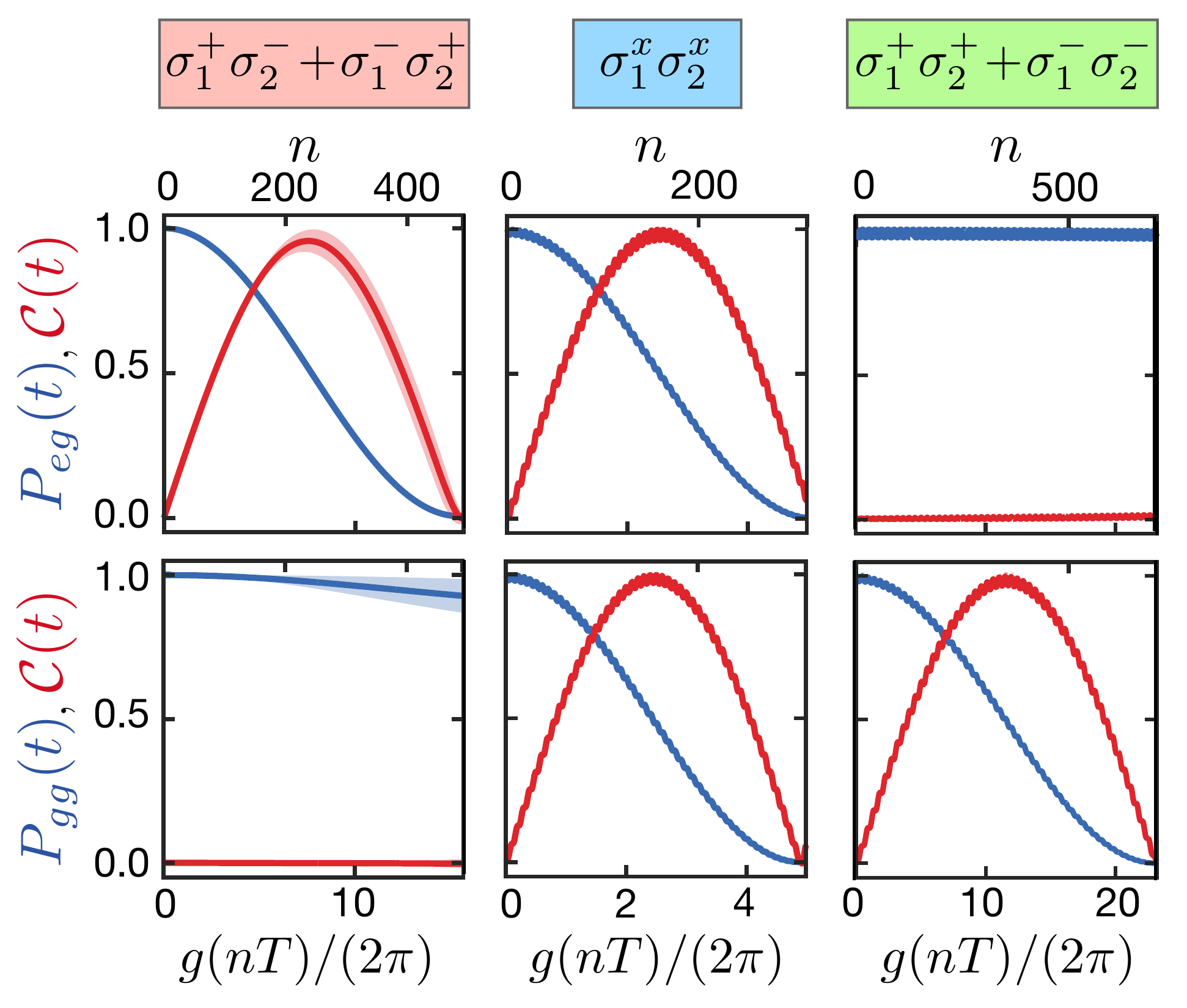}
	\caption{Numerical benchmarking of the different types of effective flip-flop (left), Ising (middle) and squeezing (right) interactions discussed in Sec.~\ref{subsec:EffectiveSS}. The upper (lower) panels show the exact time evolution of initial state $|eg\rangle$ ($|gg\rangle$) in terms of the  population $P_{eg}$ ($P_{gg}$) in blue and the concurrence $\mathcal{C}$ of the reduced TLSs state in red. The solid lines and shaded areas represent the mean values and standard deviation of these quantities, as obtained from averaging over $500$ realizations of static fluctuations with strength $\sigma=0.3 g$. In all simulations we assume an $X_{T/2}X_{T}$ sequence with a pulse duration of $\tau_\pi=0.01\tau$. The other relevant parameters for the left panel are $m=2$ and $T=0.2g^{-1}$. For the Ising and the squeezing interactions these are $m=0$, $T=0.1g^{-1}$ and $\Delta_{\rm eff}=10g$, and $m=3$, $T=0.2g^{-1}$, and $\Delta_{\rm eff}=-2.16g$, respectively. The relevant detunings $\Delta$ are indicated by the respective colored bars in Fig.~\ref{Fig:4}(b).}
	\label{Fig:5}
\end{figure}

\subsubsection{Numerical benchmarks}
In Fig.~\ref{Fig:5}, we benchmark the three different types of spin-spin interactions discussed above by exact numerical simulations of Hamiltonian~\eqref{eq:TCHamil}, interrupted by finite-width $\pi$-pulses of duration $\tau_\pi=10^{-2}\tau$. For concreteness, we consider the time evolution of two distinct initial states, $|eg\rangle$ and $|gg\rangle$, and plot the resulting probabilities $P_{eg}(t)$ and $P_{gg}(t)$ together with the concurrence $\mathcal{C}(t)$ of the reduced state of the two TLSs as a function of time. 
Each of the columns shows the results of one type of spin-spin interaction, with the parameters of the XX sequence adjusted accordingly (see details in the figure caption). In all cases, we observe an excellent agreement between the full dynamics and the effective model, even under conditions where the noise strength is comparable to the bare coupling strength $g$ and therefore exceeds by far the effective interaction strength, $\sigma\gg J$. Note, that in these examples we have assumed a rather large number of pulses, as indicated by the number of periodic repetitions, $n=N_\pi/2$. This could be relaxed, for example, in the case of flip-flop interactions, by choosing higher values of $m$ to increase the period $T$ while keeping $\Delta$ (and thus $J$) constant.

\ia{Finally, let us emphasize that the different interaction engineering strategies presented here are not exhaustive and can be combined, interchanged and optimized, depending on the application. For example, by using a XXYY sequence, the flip-flop interaction can also be realized via a detuning first-order effective JC interaction with a coupling strength $J_{\rm eff}\approx g_{\rm eff}^2/\Delta_{\rm eff}$, which can be significantly larger than the second-order coupling $J$ assumed in the example above.}

\section{Imperfections}
After having explored various different ways to employ pulsed DD schemes for engineering protected cavity QED Hamiltonians, let us now take a closer look at the potential gain one can achieve with this strategy under realistic conditions. 
To do so, we must, in addition to frequency fluctuations and pulse imperfections, also include the Markovian decay of the cavity mode with rate $\kappa$, which we have omitted so far from our analysis.

\subsection{Pulse errors and noise}\label{subsec:ErrorModel}
As a first step, however, we address purely coherent errors, which arise, for example, from nonideal pulses or from an incomplete suppression of noise. To treat such errors in a systematic manner, we evaluate the correction Hamiltonian $H_{\rm corr}$ introduced in Eq.~\eqref{eq:effHamil1} up to second order in the coupling $g$ and write the result as 
\begin{eqnarray}\label{eq:CorrHamil_expanded_main}
\tilde H_{\rm corr}&\simeq & \tilde H_{\rm corr}^{(0)}+\sum_j (\xi_jT)\, \tilde H_{\rm corr}^{(1)} \\ &+&\sum_j(\xi_jT)^2\, \tilde H_{\rm corr}^{(2,1)}+ (\xi_1\xi_2T^2)\, \tilde H_{\rm corr}^{(2,2)}. \nonumber
\end{eqnarray}
The full expressions for each of these contributions are presented together with the derivation of all the results in this section in Appendix~\ref{App:FullEff}. 

\subsubsection{Finite pulse spacing} 
The first term in Eq.~\eqref{eq:CorrHamil_expanded_main} is independent of the noise and represents, for example, residual errors arising from a finite width or spacing between the pulses. From an explicit evaluation of this term for XXYY and XY8 sequences, we find that the main contribution arises from a nonvanishing spacing between the pulses $\sim gT$ and results in an effective Hamiltonian of the form
\begin{equation}\label{eq:Hcorr_gT}
H_{\rm corr}^{(0)}\approx \frac{g^2T}{4}G_{2,z}^{(0)}\left(a^{\dag 2}+ a^2\right) \sigma_z.
\end{equation} 
The numerical prefactor, which is $G_{2,z}^{(0)}=0.25$ for XXYY$_{m=0}$ and $G_{2,z}^{(0)}\approx 0.20$ for XY8$_{m=2}$, shows no strong dependence on the applied sequence. For the latter case, the resulting transfer error is given by the second term in Eq.~\eqref{eq:XY8Error}.  Due to a larger effective coupling strength and shorter period, an even lower error can be achieved with the XXYY$_{m=0}$ sequence, as confirmed by the inset of Fig.~\ref{Fig:3}(b) for $\sigma\rightarrow 0$. 

Note that in the absence of noise and up to the order considered in the expansion, a finite width of pulses $\tau_{\pi}$ has no direct influence on the fidelity. However, this advantage is lost when larger values of $\sigma$ are taken into account, where we find that XXYY$_{m=0}$ with $\tau_\pi>0$ performs worse than the corresponding XY8$_{m=2}$ sequence. For the implementation of off-resonant spin-spin interactions, $\tau_\pi$ must be short compared to $\sigma^{-1}$ and $g^{-1}$, but the pulses can be slow compared to $\Delta^{-1}$. For example, this is the case for most of the results presented in Fig.~\ref{Fig:6}. Thus, even in this limit, the implementation of the decoupling scheme remains experimentally feasible.

\subsubsection{Noise cancellation} 
The second term in Eq.~\eqref{eq:CorrHamil_expanded_main} captures first-order corrections from the noise. For a resonant JC interaction, its main effect is a random modulation of the effective coupling strength,
\begin{equation}\label{eq:g_mod}
g_{\rm eff}(\xi)\simeq g \left( \eta + \xi T O_1\right),
\end{equation} 
where $O_1$ is a sequence-dependent numerical factor. On average, this correction induces a transfer error $\sim (\sigma \tau)^2$, which, for the XY8$_{m=2}$ sequence, is captured by the first term in Eq.~\eqref{eq:XY8Error}. For the XXYY$_{m=0}$ sequence we find that $O_1=0$ and this contribution vanishes. Therefore, we must evaluate the higher-order corrections in the second line of Eq.~\eqref{eq:CorrHamil_expanded_main}, which scale as $\sim \xi^2$. These terms result in a transfer error of
\begin{eqnarray}\label{eq:Error_t_XXYY}
\mathcal{E}_{\rm t} \approx   2.61 \times\frac{(2\sigma)^4\tau^2}{\eta^2g^2}{\Gamma_{\rm pw}}^2,
\end{eqnarray}
which now scales as $\sigma^4$. For the XXYY$_{m=0}$ sequence with $\tau_\pi/\tau=0.1$ the numerical factor is $\Gamma_{\rm pw}\approx 0.4$, but it vanishes for $\tau_\pi=0$. In this limit the error is determined by even high-order processes, which are no longer included in our error model. 

For the realization of cavity-mediated spin-spin interactions, we can choose $m=4,8,12...$ such that also for the XY8$_{m}$ sequence, the first-order coupling correction vanishes, $O_1=0$. The main correction then arises from a modulation of the exchange coupling term, $J \rightarrow J+ \delta J$, where  
\begin{eqnarray}
\delta J\approx -\frac{JT^2}{4^35^2}(\xi_1-\xi_2)^2.
\end{eqnarray}
Interestingly, also in this case the lowest-order correction is quadratic in the $\xi_i$, which for uncorrelated noise leads to an entanglement error of 
\begin{eqnarray}\label{eq:entanglement_error}
\mathcal{E}_{\rm e}\approx \frac{12\pi^2}{5^4} (\sigma \tau)^4.
\end{eqnarray}

\subsection{Cavity decay}\label{subsec:Decay}
To account for incoherent losses of the cavity mode, the system must be described by a master equation of the form 
\begin{equation}\label{eq:fullME}
\dot \rho = -i [H(t), \rho] + \frac{\kappa}{2}\left(2 a\rho a^\dag -a^\dag a\rho - \rho a^\dag a\right),   
\end{equation} 
where $\rho$ is the density operator of the full system and $\kappa$ is the cavity decay rate. 

\subsubsection{Strong coupling regime} 
In the case of resonant interactions, we are usually interested in the regime, $g_{\rm eff} \gg\kappa$, where a coherent exchange of excitations between the TLS and the cavity can take place. This condition implies $\kappa T\ll 1$ and we can simply replace $H(t)$ by $\tilde H_{\rm eff}^{(1)}(t)$ in Eq.~\eqref{eq:fullME} to evaluate the combined system dynamics with a lossy cavity. The DD pulses then suppress the effect of the noise approximately as $\sigma \rightarrow \sigma/N_\pi$, while having no influence on the Markovian decay of the cavity mode. Thus, we obtain the relaxed strong-coupling condition,
\begin{equation}
g_{\rm eff}> \sigma/N_\pi,\kappa, 
\end{equation}
as long as a sufficiently large number of pulses is applied.  

\subsubsection{Weak coupling regime} 
In the weak-coupling or far-detuned regime, the cavity is only virtually populated and its dynamics can be adiabatically eliminated. For the bare JC model, this procedure predicts a decay of the excited state of the TLS with rate
$\gamma_{0}  \simeq  g^2 \kappa/(\Delta^2+\kappa^2/4).$
In Appendix~\ref{app:CavityDecay} we extend this analysis and derive an equivalent effective rate for arbitrary DD pulses. In the limit of large detuning $\Delta \gg T^{-1},\kappa$ the resulting decay rates for the ground and excited state simplify to 
\begin{equation}
\gamma^e_{\rm eff}\simeq \gamma^g_{\rm eff}\simeq \frac{g^2}{2\Delta^2} \kappa.
\end{equation} 
This is approximately half of $\gamma_0$, the rate of the excited state in an undriven TLS. However, in the modulated system, both states are affected equally and the resulting decoherence rate of a superposition state remains $\gamma_0$.

In the opposite limit $\Delta T\sim 1$, and for sequences with a nonvanishing first-order JC-like coupling, $\eta\neq 0$, our  analysis predicts the decay of a TLS initialized in state $|e\rangle$ with a rate  
\begin{equation}
\gamma^e_{\rm eff}  \simeq  \frac{g_{\rm eff}^2 \kappa}{\Delta_{\rm eff}^2+\kappa^2/4}.
\end{equation} 
This rate is as expected for a JC-model with effective parameters, $g_{\rm eff}$, $\Delta_{\rm eff}$, and $\kappa$. Note, however, that also in this regime the ground state decays with a small, but nonvanishing rate $\gamma_{\rm eff}^g\approx \gamma_0$.

For general detunings, the  expressions for $\gamma^{e/g}_{\rm eff}$ are more involved, but can be evaluated for arbitrary pulse sequences, as described in Appendix~\ref{app:CavityDecay}. A typical behavior for the XY8 sequence is shown in Fig.~\ref{Fig:App2}. Overall, we find that away from individual resonances the simple scaling $\gamma^{e/g}_{\rm eff}\approx \gamma_0$ provides a useful general estimate for the effective decay rates.

\subsection{Enhanced cooperativity under pulsed DD}\label{subsec:Coop}
Finally, let us return to the application of generating entanglement via a cavity-mediated flip-flop interaction in the far-detuned regime. For the bare JC model, but taking both frequency noise and cavity decay into account, the resulting entanglement fidelity is approximately given by 
\begin{equation}\label{eq:ErrorModelFree}
\langle\mathcal{F}_{\rm e}\rangle\approx 1-\frac{4}{\pi^2}\left(\frac{ T_{\rm e}}{T_2^*}\right)^2 - \gamma_{0}T_{\rm e} 
\end{equation}
with $T_2^*=\sqrt{2}/\sigma$. This expression reaches a maximum for $\Delta_{\rm opt}=(\pi g^4  \kappa (T_2^*)^2/2)^{1/3}$, where 
\begin{equation}\label{eq:MinErrorFree}
\langle\mathcal{F}_{\rm e}\rangle\approx 1- \frac{3}{8}\left(\frac{\pi}{\mathcal{C}}\right)^{2/3}.
\end{equation}
Therefore, in this detuned regime, the maximal fidelity depends only on a single parameter, namely the cooperativity
\begin{equation}
\mathcal{C} = \frac{g^2}{\sigma \kappa}.
\end{equation}
 A similar scaling as in Eq.~\eqref{eq:MinErrorFree} is also found for several other applications in cavity QED, where larger values of $\kappa$ can be compensated by correspondingly lower dephasing rates. 

\begin{figure}
\includegraphics[width=\columnwidth]{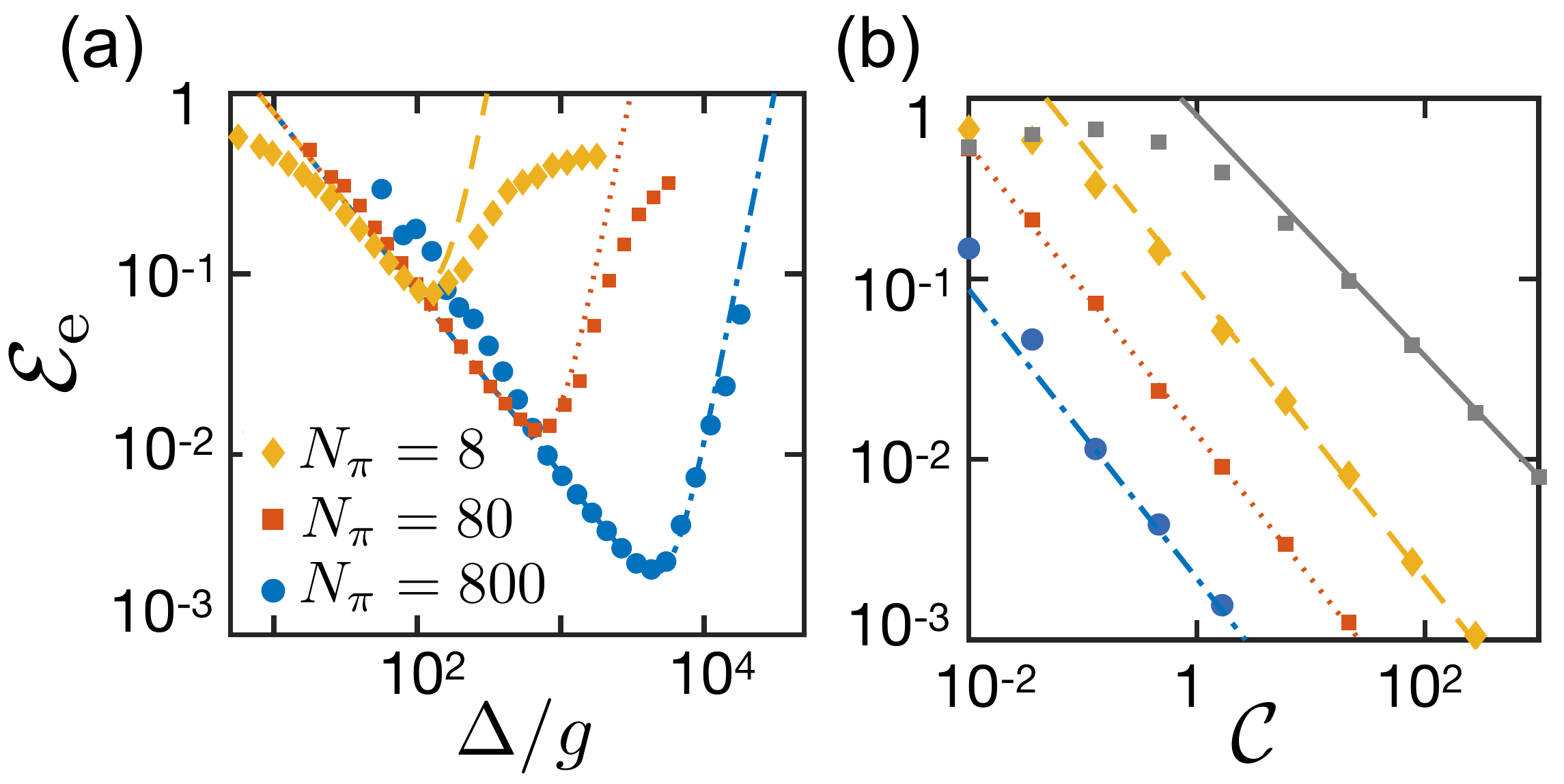}
	\caption{(a) Plot of the entanglement error $\mathcal{E}_{\rm e}=1-\langle \mathcal{F}_{\rm e}\rangle$ as a function of the detuning $\Delta$ and for different numbers of pulses $N_\pi$. For this simulation, we consider the implementation of a flip-flop interaction with an XY8 sequence ($t_1=0.5\tau$ and $\tau_\pi=0.01\tau$) and for $\kappa=10 g$ and $\sigma=0.1 g$ ($\mathcal{C}=1$). The averages are taken over 500 realizations of static fluctuations. The diamond, square and round markers are the result of numerical simulations, and the dashed, dotted and dashed-dotted lines represent the prediction of Eq.~\eqref{eq:ErrorModelXY8} for $N_\pi=8, 80$ and $800$, respectively. Note that we limit the possible values of the detuning to $\Delta=g\sqrt{m N_\pi }$ with $m=4,8,12...$. (b) The minimum achievable error $\mathcal{E}_{\rm min}$ is plotted versus the cooperativity $\mathcal{C}$ for different number of pulses $N_\pi$. The markers represent the results of numerical simulations with $\Delta\approx \Delta_{\rm opt}$, while the lines indicate the scaling given in  Eq.~\eqref{eq:MinErrorXY8}. The gray square markers and the solid line represent exact results and the scaling in Eq.~\eqref{eq:MinErrorFree} for the pulse-free case.} 
	\label{Fig:6}
\end{figure}

In Fig.~\ref{Fig:6} we simulate the same entangling gate under the influence of DD pulses. Specifically, for these simulations we consider the XY8$_{m}$ sequence, which is very robust with respect to pulse imperfections and thus well-suited for implementing DD schemes with a very large number of pulses~\cite{Wang:2016}. For detunings $\Delta= m\times 2\pi /T$ with $m=4,8,...$ the first-order coupling vanishes, $g_{\rm eff}=0$, and the effective system evolution is well-described by the flip-flop Hamiltonian in Eq.~\eqref{eq:flipflopHamil} with $J=g_1g_2/\Delta$.
To ensure decoupling during all periods, we choose the final time $t\simeq T_{\rm e}$ as a multiple of the period $T$, and thus set $\Delta=g\sqrt{m N_\pi}$. For a fixed number of pulses $N_\pi$, we vary $\Delta$ by changing the value of $m=4,8,12,...$.

Under these conditions,  our error analysis from above predicts an average entangling fidelity of 
\begin{equation}\label{eq:ErrorModelXY8}
\langle\mathcal{F}_{\rm e}\rangle\approx 1-\frac{3\pi^2}{5^4N_\pi^4}\left( \frac{T_{\rm e}}{T_2^*}\right)^4 - \gamma_{0}T_{\rm e},
\end{equation}
where, as discussed above, we can use $\gamma_{0}=g^2\kappa/\Delta^2$ as the approximate decay rate for large detunings.
This expression shows that the DD pulses not only increase the coherence time, but also change the scaling of the error for static noise. 
Therefore, the fidelity is optimized at a different detuning $\Delta_{\rm opt}=2.13 \,  (g^8 N_\pi^4  \kappa /\sigma^4)^{1/5}$, where it reaches a maximal value of 
\begin{equation}\label{eq:MinErrorXY8}
\langle\mathcal{F}_{\rm e}\rangle\approx 1- 0.46\,\Big(\frac{1 }{N_\pi \mathcal{C}}\Big)^{4/5},
\end{equation}
predicting an almost linear gain with the number of applied $\pi$-pulses.

To confirm our analytical estimates, in Fig.~\ref{Fig:6}(a) we show the excellent correspondence between Eq.~\eqref{eq:ErrorModelXY8} and the entanglement fidelity obtained from exact numerical simulation over a wide range of parameters, as long as overall error is small enough, $\mathcal{E}_{\rm e}\lesssim 0.1$.  In Fig.~\ref{Fig:6}(b) we also compare the optimal fidelities given in Eq.~\eqref{eq:MinErrorFree} and Eq.~\eqref{eq:MinErrorXY8} with the corresponding numerically optimized results for different values of the cooperativity $\mathcal{C}$. We vary the latter by changing $\sigma$ while maintaining a fixed $\kappa=10g$. Again, the exact results follow very accurately the predicted trends and confirm the boost of the effective cooperativity by several orders of magnitude. \ia{The same enhancement will also benefit many other applications, for example, cavity-mediated spin squeezing, where the minimal spin-squeezing parameter is predicted to scale as $\xi^2_{\rm squeez} \sim 1/\sqrt{N_\pi \mathcal{C}}$~\cite{Bennett:2013}.}

Note that all analytical and numerical predictions in this work assume that any residual incoherent decay of the TLS with rate $T_1^{-1}$ is negligible on the timescales of interest. Such a Markovian decay is not affected by DD and will contribute a trivial error $O(T_{\rm e}/T_1)$ to any coherent operation.

\section{Conclusion}
In summary, we have proposed a general pulsed DD strategy for protecting cavity-QED systems against quasi-static frequency fluctuations. Our analysis revealed that this approach not only suppresses the effects of noise but also enables the engineering and modulation of different types of interactions by simply adjusting the pulse parameters. Furthermore, we provided a comprehensive analysis of the effective interactions and residual errors that arise for a given DD sequence, facilitating the optimization of this technique for specific experimental setups.

As a relevant application, we demonstrated how cavity-mediated entanglement operations can be systematically enhanced by increasing the number of applied $\pi$-rotations using the experimentally robust XY8 sequence. These findings are particularly relevant for solid-state cavity-QED experiments with spin qubits~\cite{Wang2023,Dijkema:2025,Mi:2018,Samkharadze:2018,Landig:2018} or rare-earth dopants~\cite{Zhong:2019,Kindem:2020,Ourari:2023,Gritsch:2025}, where frequency inhomogeneities and slow frequency drifts present common experimental challenges. However, these techniques also offer a robust and versatile approach for engineering effective light-matter and spin-boson interactions in a wide range of other settings.

\section{Acknowledgements}
We acknowledge support from the Swiss National Science Foundation through Project Nr. CRSII 222812/1 and from the European Union’s Horizon Europe research and innovation program under grant agreement No 101114305 (``MILLENION-SGA1" EU Project). P.~B. acknowledges funding from the France 2030 plan under the ANR-22-PETQ-0003 grant. This research is part of the Munich Quantum Valley, which is supported by the Bavarian state government with funds from the Hightech Agenda Bayern Plus.


\appendix

\setcounter{figure}{0}
\renewcommand{\figurename}{Fig.}
\renewcommand{\thefigure}{A\arabic{figure}}

\section{Errors in the JC state transfer}\label{app:TranferProb}
For all our numerical simulations and analytic estimates, we consider the limit of a purely static noise, $\tau_c/\tau\rightarrow 0$. In this appendix, we quantify how small static shifts in the experimental parameters affect the JC state transfer discussed in Sec.~\ref{subsec:vaccumRabi}. We consider deviations from the ideal JC Hamiltonian $H=g_0 (\sigma_+a+\sigma_-a^\dagger)$, modeled by the Hamiltonian
\begin{equation}
H=\frac{\xi}{2}\sigma_z  +  ( g\sigma_+ a+ {g}^*\sigma_- a^\dagger ),
\end{equation}
where $g=g_0(1+\epsilon)$. Here $\xi$ accounts for frequency fluctuations of the TLS, and $\epsilon\in\mathbb{C}$ for deviations in the value of the coupling parameter $g_0$. If we initialize the system in state $|e,0\rangle$, the transfer fidelity after time $T_{\rm t}=\pi/(2g_0)$ is
\begin{equation}
\mathcal{F}_{\rm t}(T_{\rm t})=\frac{|g|^2}{\tilde g^2}\sin^2\Big({\frac{\pi}{2}\frac{\tilde{g}}{g_0}}\Big),
\end{equation}
where $\tilde{g}=\sqrt{|g|^2+\xi^2/4}$. Assuming that the deviations are small, $\xi\ll g_0$ and $|\epsilon| \ll 1$, we obtain
\begin{equation}
\tilde{g}/g_0\approx 1+{\rm Re}(\epsilon)+\frac{1}{2}\left[|\epsilon|^2+\frac{\xi^2}{4g_0^2}\right],
\end{equation}
and the transfer fidelity can be approximated by
\begin{equation}\label{eq:JCSTFidelity}
\mathcal{F}_{\rm t}(T_{\rm t})\approx 1-\frac{\xi^2}{4g_0^2}-\frac{\pi^2}{4}\left[{\rm Re}(\epsilon)+\frac{|\epsilon|^2}{2}\right]^2.
\end{equation}
When we restrict ourselves to frequency fluctuations with strength $\langle\xi^2\rangle=\sigma^2$ only, Eq.~\eqref{eq:JCSTFidelity} reduces to Eq.~\eqref{eq:bare_fidelity} of the main text. The other contributions $\sim \epsilon$ are relevant to characterize other types of imperfections considered in Appendix~\ref{app:ErrorEstimates}.

\section{Derivation of the full effective Hamiltonian}\label{App:FullEff}
\begin{figure}
\includegraphics[width=\columnwidth]{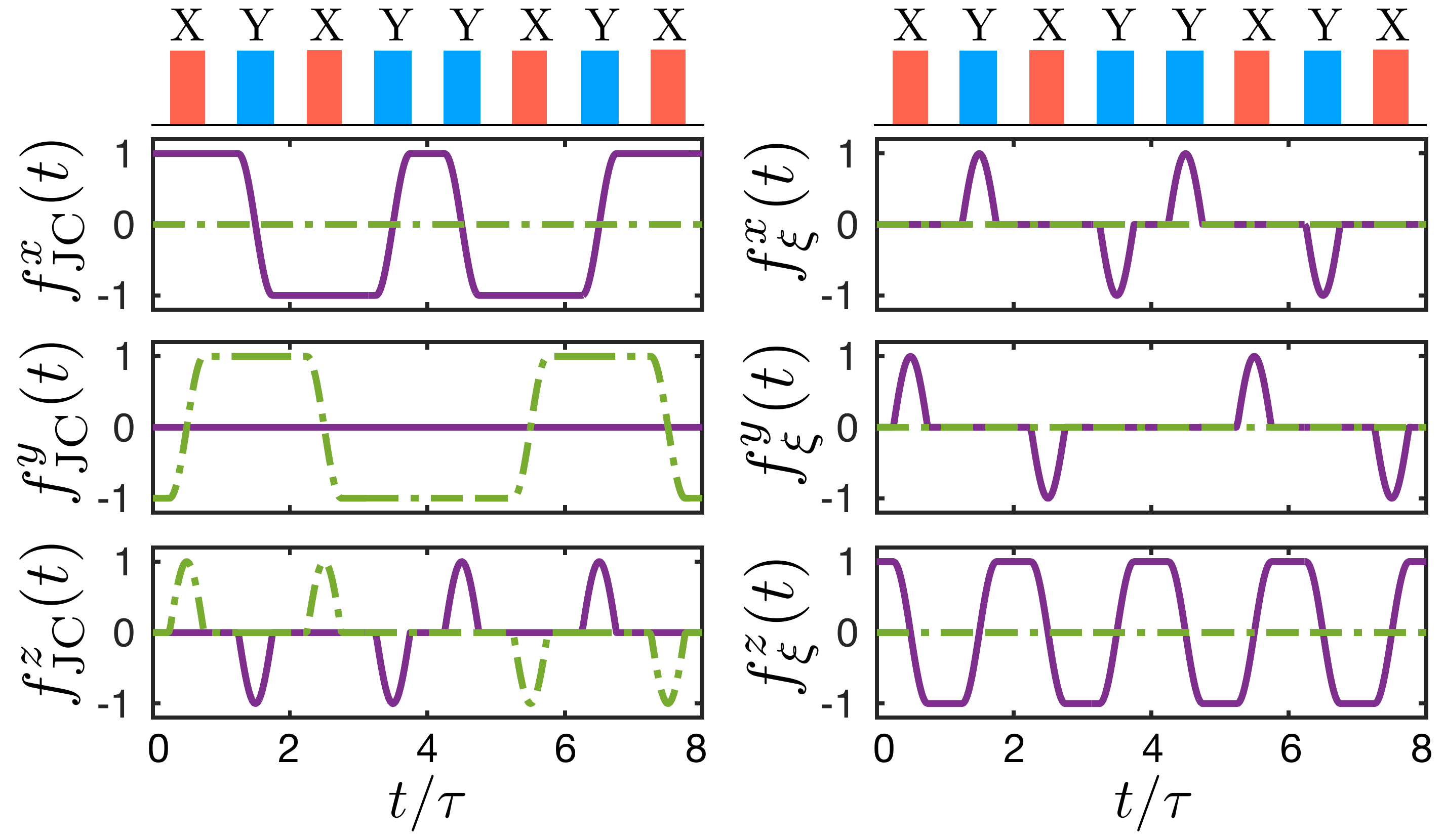}
	\caption{Illustration of the time-dependence of the modulation functions $\vec{f}_{\rm JC}(t)$ (left plots) and $\vec{f}_{\xi}(t)$ (right plots) for an XY8 sequence with pulses of width $\tau_\pi=0.5 \tau$. The solid and dashed lines indicate the real and imaginary parts, respectively. In the limit $\tau_\pi\rightarrow 0$, the functions $f_{\rm JC}^z$ and  $f_{\rm \xi}^{x,y}$ vanish identically.} 
	\label{Fig:A1}
\end{figure}
In this appendix we summarize the details of the derivation of the full second-order Hamiltonian given in Eq.~\eqref{eq:effHamil0}, which contains the targeted effective first-order and second-order interactions, as well as additional correction terms. For the derivation of this effective model we start from the cavity QED Hamiltonian in Eq.~\eqref{eq:TCHamil} in the toggling frame, where 
\begin{equation}\label{eq:FullHamil}
\begin{split}
\tilde H(t)=& \sum_{j=1}^{2}\frac{\xi_j}{2}\vec{f}_{\xi}(t)\cdot\vec{\sigma}_j \\
+&\sum_{j=1}^{2}\frac{g_j}{2} ( \vec{f}_{\rm JC}(t)\cdot\vec{\sigma}_j\, a^\dagger  e^{i\Delta t}+{\rm H.c.} ).
\end{split} 
\end{equation}
Note that in contrast to the analysis in the main text, here we are more general and include pulses of finite duration. In this case the modulation functions $\vec{f}_\xi(t)$ and $\vec{f}_{\rm JC}(t)$ are defined by
\begin{equation}
\begin{split} 
U_\pi^\dagger(t) \sigma_j^z U_\pi (t)=& \vec{f}_\xi(t)\cdot \vec{\sigma}_j,  \\
U_\pi^\dagger(t) \sigma_j^- U_\pi (t)=& \vec{f}_{\rm JC}(t)\cdot \vec{\sigma}_j.  \\
\end{split} 
\end{equation}
Thus, in this general case, the transformation to the toggling frame mixes the different Pauli operators. For illustration, Fig.~\ref{Fig:A1} shows the form of these functions for the XY8 sequence and for pulses with a deliberately long duration of $\tau_\pi=0.5\tau$. For instantaneous pulses, these functions simplify to $\vec{f}_{\rm JC}(t)=(f_x(t), -i f_y(t),0)$ and $\vec{f}_{\xi}(t)=(0,0,f_z(t))$, where the $f_k(t)$ behave according to Eq.~\eqref{eq:fliprules}.

In a second step, we proceed by implementing another unitary transformation into an interaction picture with respect to the noise term $\tilde H_\xi(t)=\sum_j \xi_j f_\xi^z(t)\sigma_j^z/2$. The purpose of this transformation is not immediately obvious, but, as we show below, it allows us to capture relevant corrections terms $\sim g^2\xi^2$, which we would not obtain from a direct second-order expansion. In this new frame, Eq.~\eqref{eq:FullHamil} transforms into 
\begin{equation}\label{eq:Htilde_int_xi}
\begin{split}
\tilde H(t)=& \sum_{j=1}^{2}\frac{\xi_j}{2}\vec{F}_{\xi,j}(t)\cdot\vec{\sigma}_j \\
+&\sum_{j=1}^{2}\frac{g_j}{2} \left( \vec{F}_{\rm JC,j}(t)\cdot\vec{\sigma}_j\, a^\dagger +{\rm H.c.} \right),
\end{split} 
\end{equation}
where the updated modulation functions now depend on $\varphi_j(t)= \xi_j  \int_0^t ds f_z^\xi(s)$ as
\begin{eqnarray}
F_{{\rm JC},j}^x(t)&=&[f_{{\rm JC},j}^x(t)\cos{\varphi_j(t)}+f_{{\rm JC},j}^y(t)\sin{\varphi_j(t)}]e^{i\Delta t},  \nonumber  \\
F_{{\rm JC},j}^y(t)&=&[f_{{\rm JC},j}^y(t)\cos{\varphi_j(t)}-f_{{\rm JC},j}^x(t)\sin{\varphi_j(t)}]e^{i\Delta t},\nonumber\\
F_{\xi,j}^x(t)&=&f_{\xi,j}^x(t)\cos{\varphi_j(t)}+f_{\xi,j}^y(t)\sin{\varphi_j(t)},  \nonumber \\
F_{\xi,j}^y(t)&=&f_{\xi,j}^y(t)\cos{\varphi_j(t)}-f_{\xi,j}^x(t)\sin{\varphi_j(t)},\nonumber
\end{eqnarray}
and $F_{{\rm JC},j}^z(t)=f_{{\rm JC},j}^z(t)$, and $F_{{\xi},j}^z(t)=0$. Note that, in this frame, the first term in Eq.~\eqref{eq:Htilde_int_xi} accounts exclusively for effects related to a finite pulse width and vanishes for instantaneous pulses.

At this stage we perform a Magnus expansion for the evolution operator during the $n$-th time interval $[nT,(n+1)T]$. More precisely, we write
\begin{equation}
\mathcal{T}e^{-i\int_{nT}^{(n+1)T} dt  \tilde H(t)} = e^{-i \sum_{k=1}^\infty H_{\rm eff}^{(k)}(nT) T},
\end{equation}
where $\mathcal{T}$ denotes the time-ordered exponential. By truncating this expansion after the second-order and by making use of the identity $[\vec a \cdot \vec \sigma_j, \vec b \cdot \vec \sigma_k] = 2i\delta_{j,k}\,(\vec a \times \vec b) \cdot \vec \sigma_k$, we finally obtain the effective Hamiltonian
\begin{eqnarray}\label{eq:effHamil}
\tilde H_{\rm eff}(nT) &=& 
\sum_j\frac{\xi_j}{2} \vec{\mathcal{W}}_j\cdot\vec \sigma_j\\
 &+&\sum_j\frac{g_j}{2}\big\{\vec{\mathcal{O}}_j a^\dagger e^{i\Delta_{\rm eff} nT}+{\rm H.c.}\big\} \cdot\vec{\sigma}_j \nonumber\\
 &+&\sum_j\frac{g_j^2 T}{4}\big\{\vec{\mathcal{G}}_{2,j}(a^\dagger)^2 e^{i2\Delta_{\rm eff} nT} + {\rm H.c.}\big\} \cdot \vec{\sigma}_j \nonumber \\&+& \sum_j\frac{g_j^2 T}{4} (2a^\dagger a+1) {\rm Re}\big(\vec{\mathcal{G}}_{3,j}\big)  \cdot \vec{\sigma}_j  \nonumber \\&+&   
  \sum_{\{u,v\}=\{x,y,z\}} \frac{g_1g_2 T}{2}{\rm Im}(\mathcal{G}_{uv})\, \sigma_1^u \sigma_2^v. \nonumber
\end{eqnarray}
Here we introduced the sets of dimensionless parameters $\vec{\mathcal{W}}_j=\vec{\Gamma}_j+\frac{\xi_jT}{2}\vec{\mathcal{G}}_{0,j}$, and $\vec{\mathcal{O}}_j=\vec{O}_j+\xi_jT\vec{\mathcal{G}}_{1,j}$, which in turn 
are given in terms of the pulse-dependent integrals
\begin{eqnarray}
\vec{\Gamma}_j=\frac{1}{T}\int_0^T dt \, \vec{F}_{\xi,j}(t), \\
\vec{O}_j=\frac{1}{T}\int_0^T  dt \,\vec{F}_{{\rm JC},j}(t) ,
\end{eqnarray}
and
\begin{eqnarray}
\vec{\mathcal{G}}_{0,j}&=&\frac{1}{T^2}  \iint \vec{F}_{\xi,j}(t)\times\vec{F}_{\xi,j}(s), \\
\vec{\mathcal{G}}_{1,j}&=&\frac{1}{T^2}\iint \vec{F}_{{\rm JC},j}(t)\times\vec{F}_{\xi,j}(s) ,\\
\vec{\mathcal{G}}_{2,j}&=&\frac{1}{T^2} \iint \vec{F}_{{\rm JC},j}(t)\times\vec{F}_{{\rm JC},j}(s) , \\ 
\vec{\mathcal{G}}_{3,j}&=& \frac{1}{T^2} \iint \vec{F}^*_{{\rm JC},j}(t)\times\vec{F}_{{\rm JC},j}(s) , \\
\mathcal{G}_{uv}&=&\frac{1}{2T^2}\iint  \, [ \hat{u}\cdot\vec{F}_{{\rm JC},1}^*(t)][\hat{v}\cdot \vec{F}_{{\rm JC},2}(s)], \nonumber \\
&+&\frac{1}{2T^2}\iint  \, [ \hat{u}\cdot\vec{F}_{{\rm JC},2}^*(t)][\hat{v}\cdot \vec{F}_{{\rm JC},1}(s)],
\end{eqnarray}
where $\iint\equiv \int_{0}^{T} dt \int_{0}^tds$ for clarity. It is noteworthy that the numerical value of these integrals depends only on the form of the pulse sequence, and the dimensionless parameters $\Delta T$ and $\xi_j T$. 

In the form given in Eq.~\eqref{eq:effHamil}, the effective Hamiltonian $H_{\rm eff}$ contains all interaction terms up to order $(gT)^2$, but in view of the interaction representation assumed in Eq.~\eqref{eq:Htilde_int_xi}, it still contains arbitrary orders of $\xi_j$. Therefore, in a final step we expand $\vec{F}_{\xi,j}(t)$ and $\vec{F}_{{\rm JC},j}(t)$ up to second order in $\xi_j$ (assuming $\varphi_j(t)\ll1$), leading to the decomposition of the integrals $\vec{\Gamma}_j, \vec{O}_j$ and $\vec{G}_{k,j}$ in the fashion
\begin{eqnarray}
\vec{\mathcal{G}}_{k,j}=\vec{\mathcal{G
}}_{k}^{(0)}+(\xi_j T)\vec{\mathcal{G}}_{k}^{(1)}-\frac{1}{2}(\xi_j T)^2\vec{\mathcal{G}}_{k}^{(2)},
\end{eqnarray}
and the decomposition of $\mathcal{G}_{uv}$ as
\begin{eqnarray}
\mathcal{G}_{uv}&=&\mathcal{G}_{uv}^{(0)}+\frac{1}{2}(\xi_1+\xi_2)T \mathcal{G}_{uv}^{(1)}  \\&-& \frac{1}{4}(\xi^2_1+\xi^2_2)T^2 \mathcal{G}_{uv}^{(2,1)} - \frac{1}{4}(\xi_1\xi_2)T^2 \mathcal{G}_{uv}^{(2,2)}.\nonumber
\end{eqnarray}
Hamiltonian~\eqref{eq:effHamil} can then be reorganized accordingly as
\begin{eqnarray}\label{eq:CorrHamil_expanded}
\tilde H_{\rm eff}&\approx& \tilde H_{\rm sc}^{(1)}+\tilde H_{\rm ss}^{(2)}+ \tilde H_{\rm corr}^{(0)}+\sum_j (\xi_jT)\, \tilde H_{\rm corr}^{(1)}\nonumber  \\ &+&\sum_j(\xi_jT)^2\, \tilde H_{\rm corr}^{(2,1)}+ (\xi_1\xi_2T^2)\, \tilde H_{\rm corr}^{(2,2)}, 
\end{eqnarray}
where $\tilde H^{(1)}_{\rm sc}$ and $\tilde H^{(2)}_{\rm ss}$ are the interaction terms given in Eqs.~\eqref{eq:1stHamil} and \eqref{eq:HamilManyPulse} in the main text with $\eta_x\equiv O_x^{(0)}$, $\eta_y \equiv iO^{(0)}_y$ and $r_{uv}\equiv-\Delta T \mathcal{G}_{uv}^{(0)}$. All the remaining terms represent noise-induced and other pulse-related corrections. Note that while avoiding a full 4th-order Magnus expansion, our derivation still accounts for relevant correction terms scaling as $\sim g^2 \xi^2$. The numerical values for the most relevant coefficients are summarized in Table \ref{table:1} for the sequences XXYY$_{m=0}$, XY8$_{m=2}$, and \ia{$X_{T/2} X_{3T/4}Y_{3T/4} Y_T$~\cite{Groszkowski:2022}, the latter denoted here as XXYY$_{\rm GKLC}$}.

\setlength{\arrayrulewidth}{0.5pt}   
\begin{table*}[t]
\centering
\begin{tabular}{ || p{3.6cm}||p{0.9cm}|p{0.9cm}|p{0.9cm}|p{0.9cm}|p{0.9cm}|p{0.9cm}|p{1.2cm}|p{0.9cm}|p{0.9cm}|}
 \hline
$\times10^{-2}$ & $|O_{x,y}^{(0)}|$ & $|\mathcal{G}_{2,z}^{(0)}|$ &  $|O_{x,y}^{(1)}|$ & $\Gamma_{x,y}^{(1)}$ &   $|\mathcal{G}_{1,z}^{(1)}|$ &   $|\mathcal{G}_{2,z}^{(1)}|$ &   ${\rm Re}(\mathcal{G}_{3,{z}}^{(1)})$ & $|O_{x,y}^{(2)}|$ & $|\mathcal{G}_{2,z}^{(2)}|$  \\ [0.5ex] 
 \hline\hline
 XXYY$_{m=0}$ ($\tau_\pi/\tau\rightarrow0$)  & 50 & 25 &   0 & 0& 0  & 0  & -1.56 &  0.26  & 0.26     \\ 
 \hline
  XXYY$_{m=0}$ ($\tau_\pi/\tau=0.1$) & 50 & 25 &    0 & 0.38& 0.19  & 0  & -1.56 &  0.26  & 0.26  \\
 \hline
  XY8$_{m=2}$ ($\tau_\pi/\tau\rightarrow0$)  & 45.02 & 20.26   & 0.77& 0& 0 & 0.69 & -0.23 &  0.05 & 0.04  \\
 \hline
  XY8$_{m=2}$ ($\tau_\pi/\tau=0.1$)  & 44.99 & 20.24  & 0.77& 0& 0 & 0.69 & -0.23 &  0.05 & 0.04 \\  
\hline
  \ia{XXYY$_{\rm GKLC}$ ($\tau_\pi/\tau\rightarrow0$) } & 50 & 0   & 18.75$^*$& 0& 0 & 6.25 & 6.25 &  7.29$^*$ & 0 \\
 \hline
 \ia{ XXYY$_{\rm GKLC}$ ($\tau_\pi/\tau=0.1$)}  & 50 & 2.50   & 16.87$^*$& 0.38& 0.07 & 5.06 & 4.77 &  5.95$^*$ & 0.17 \\ 
 \hline
\end{tabular}
\caption{Numerical values of the most relevant coefficients that determine the effective Hamiltonian given in Eq.~\eqref{eq:CorrHamil_expanded} for the sequences XXYY$_{m=0}$ and XY8$_{m=2}$ introduced in Sec.~\ref{sec:ProtectedJC}, \ia{and the sequence XXYY$_{\rm GKLC}$ introduced in Ref.~\cite{Groszkowski:2022}}. For the first two sequences, $t_1=\tau/2$. For all sequences, the parameters are evaluated in the limit of infinitely short pulses, $\tau_\pi/\tau \rightarrow0 $, and for a finite pulse length of $\tau_\pi/\tau=0.1$. All other non-vanishing parameters are smaller than the ones listed in the table and do not contribute to the most relevant errors in leading order. \ia{We use ``$*$'' when $|O_{x}^{(1,2)}|$ and $|O_{y}^{(1,2)}|$ differ, in which case the larger value is given.}}
\label{table:1}
\end{table*}

\begin{table*}[t]
\centering
\begin{tabular}{ || p{3.6cm}||p{0.8cm}|p{0.8cm}|p{0.8cm}|p{0.8cm}|p{0.8cm}|p{0.8cm}|p{0.8cm}|p{0.8cm}|p{0.8cm}|p{0.8cm}|}
 \hline
$m$ & $4$ & $8$ & $12$ & $16$ &   $20$ &   $24$ &   $28$ & $32$ & $36$ & $40$  \\ [0.5ex] 
 \hline\hline
  ${\rm Im}(\mathcal{G}^{(0)}_{uu})\times(2\pi m)$    & -1 & -1 & -1& -1 & -1 & -1 & -1 & -1 & -1 &-1 \\
  \hline
  ${\rm Im}(\mathcal{G}^{(1)}_{uu})\times(2\pi m)\times 10^{2}$   & 0 & -3.98 & 0& -1.99& 0 & -1.33 & 0 &-0.99 & 0 & -0.80  \\
  \hline
  ${\rm Im}(\mathcal{G}^{(2,1)}_{uu})\times(2\pi m)\times 10^{2}$   & -0.26 & -0.10 &  -0.26 &-0.22 &-0.26 &-0.24 & -0.26 & -0.25 & -0.26 & -0.25\\
  \hline
  ${\rm Im}(\mathcal{G}^{(2,2)}_{uu})\times(2\pi m)\times 10^{2}$  & -0.11 & 0.68&  0.45&  0.56 & 0.50 & 0.54 & 0.51 & 0.53 & 0.51 & 0.53\\
 \hline
\end{tabular}
\caption{Numerical values of the relevant coefficients that determine the noise-induced errors for a cavity mediated flip-flop process. The values are given for the XY8$_{m}$ sequence discussed in Sec.~\ref{subsec:Coop}, with $t_1=\tau/2$ and pulse lengths $\tau_\pi/\tau \rightarrow 0$. Here, $u\in {x,y}$.}
\label{table:2}
\end{table*}

\section{Error estimates}\label{app:ErrorEstimates}
In its full form, the final correction Hamiltonian in Eq.~\eqref{eq:CorrHamil_expanded} is cumbersome to treat. However, as we can see from the examples in Table \ref{table:1}, not all of the correction terms are significant and in certain limits, for example, when $\tau_\pi\rightarrow 0$, many of the corrections simplify further. Therefore, Eq.~\eqref{eq:CorrHamil_expanded} is a convenient starting point to systematically evaluate the dominant effect of various different sources of imperfections for a given pulse sequence. In the following we perform such analysis in order to derive the scaling laws presented in the main text.

\subsection{Resonant JC interactions}
For the case of a single TLS, the last term in Eq.~\eqref{eq:CorrHamil_expanded} does not appear. The interaction term of interest is determined by the coefficients $O_{x,y}^{(0)}$, while the parameters $O_{z}^{(0)}$ and $\vec{\Gamma}^{(0)}$ are zero for the sequences of interest. The remaining terms depend on either $gT$ or $\xi T$ and can be suppressed by considering shorter interpulse spacings $\tau$. In the following, we consider noise-induced and other errors separately.

\subsubsection{Finite pulse spacing}
The terms in Eq.~\eqref{eq:CorrHamil_expanded} proportional to $\vec{\mathcal{G}}_{2}^{(0)}$ and $\vec{\mathcal{G}}_{3}^{(0)}$ are independent of $\xi$. They represent deviations from the ideal dynamics that arise from a finite spacing between the decoupling pulses. For all sequences of interest, $|\mathcal{G}_{2,z}^{(0)}|\gg |\mathcal{G}_{2,x,y}^{(0)}|, {\rm Re}(G_{3,u}^{(0)})$. Thus, in the absence of noise, the dominant source or errors arises from the correction given in Eq.~\eqref{eq:Hcorr_gT}. Using perturbation theory as outlined in Appendix~\ref{app:PerThe}, we can evaluate its effect on a resonant state-transfer process and obtain a transfer error scaling as 
\begin{eqnarray}
\mathcal{E}_{\rm t}\approx  1.16\times \frac{g^2T^2|\mathcal{G}_{2,z}^{(0)}|^2}{8|\eta|^2}.
\end{eqnarray}
For the XY8$_{m=2}$ sequence with $T=8\tau$ this expression simplifies to $\mathcal{E}_{\rm t}\approx 1.88 g^2\tau^2$, while for the XXYY$_{m=0}$ sequence with $T=4\tau$ we obtain
$\mathcal{E}_{\rm t}\approx 0.58 g^2\tau^2$. \ia{Compared to XXYY$_{m=0}$ and XY8$_{m=2}$, the XXYY$_{\rm GKLC}$ sequence yields smaller values of $|\mathcal{G}_{2,z}^{(0)}|$, with $|\mathcal{G}_{2,z}^{(0)}|=0$ in the limit of instantaneous pulses. In this case, the leading error is of higher order in $g\tau$ and is not captured by Eq.~\eqref{eq:CorrHamil_expanded}.}

\subsubsection{Noise cancellation}
The condition $\vec{\Gamma}^{(0)}=0$ ($\gamma_{x,y,z}=0$ in the main text) ensures that the first term in Hamiltonian~\eqref{eq:effHamil} vanishes up to lowest order in $(\xi T)$. However, at higher orders, terms in the Hamiltonian scaling as $\xi^2 T$ or $g \xi T$ still generate noise-induced deviations from the ideal dynamics. These corrections are determined by the coefficients $\vec{\Gamma}^{(1)}$ and $\vec{\mathcal{G}}^{(0)}_{0}$ and $\vec{O}^{(1)}$ and $\vec{\mathcal{G}}^{(0)}_{1}$, respectively.  In the case of the XXYY$_{m=0}$ sequence with finite pulse length $\tau_\pi=0.1\tau$, we find that $\Gamma_{\rm pw}\equiv\Gamma_{x}^{(1)}=\Gamma_{y}^{(1)}\neq 0$, while the other coefficients are zero. This implies that the dominant effect of the noise stems from a contribution of the form
\begin{equation}
\tilde H_{\rm corr}\simeq \frac{\xi^2 T}{2}\Gamma_{\rm pw} \left(\sigma_x+\sigma_y\right).
\end{equation} 
This correction induces the average transfer error given in Eq.~\eqref{eq:Error_t_XXYY}, where we again followed the general approach in Appendix~\ref{app:PerThe} and used that $\langle\xi^4\rangle=3\sigma^4$.

In the case of the XY8$_{m=2}$ sequence, we find that $\vec{O}^{(1)}\neq 0$, while all the other coefficients listed above are zero.  In particular, the two contributions proportional to ${\rm Re}(O^{(1)}_x)\equiv O_1$ and ${\rm Im}(O_y^{(1)})=-O_1$ in the second line of Eq.~\eqref{eq:effHamil} will add up to a JC interaction with a modified coupling constant given in Eq.~\eqref{eq:g_mod}. By following the general derivation in Appendix \ref{app:TranferProb}, we obtain 
\begin{eqnarray}
\mathcal{E}_{\rm t} \approx  \left(\frac{\pi\sigma T}{2|\eta|} O_1\right)^2=\left(\frac{2\sigma\tau}{3\pi}\right)^2, 
\end{eqnarray}
where for the second equality we have used that $O_1=|\eta|/(6\pi^2)$. Interestingly, for the XY8 sequence, the residual error scales as $\sigma^2$ and it also doesn't vanish in the limit of infinitely fast $\pi$-rotations. This sequence is, however, more robust with respect to other types of pulse imperfections and may still be the preferred choice under most experimental conditions. 

\ia{Finally, for the XXYY$_{\rm GKLC}$ sequence we also find that $|O_{x,y}^{(1)}|\neq 0$, but these values are much larger than those found for the XY8$_{m=2}$ sequence. This explains why the XXYY$_{\rm GKLC}$ sequence performs worse than the other sequences in the presence of strong dephasing.}

\subsection{Protected spin-spin interactions}
For the effective spin-spin interactions studied in Sec.~\ref{subsec:EffectiveSS}, the interaction terms of interest  are determined by the coefficients $O_{x,y}^{(0)}$, the effective detuning $\Delta_{\rm eff}$ and the coefficients $\mathcal{G}_{uv}^{(0)}$. For the following discussion we focus on the flip-flop interaction implemented by the same XY8 pulse sequence as described in Sec.~\ref{subsec:Coop}. For values of the detuning that fulfill $\Delta= 2\pi m/T$ with $m=4,8,12,...$, we find that $O_{x,y}^{(0)}=0$ [see Table~\ref{tab:couplings}] and ${\rm Im}(\mathcal{G}_{uu}^{(0)})=-1/(2\pi m)$, and therefore,  $J=g_1g_2/\Delta$. 

To investigate the effect of noise on the flip-flop interaction, we evaluate all parameters $\vec{\mathcal{W}}_j$, $\vec{\mathcal{O}}_j$, $\vec{\mathcal{G}}_{j,k}$ and $\mathcal{G}_{uv}$, and find that for the values of the detuning specified above the only non-zero parameters are $\mathcal{G}_{uu}^{(1)},\mathcal{G}_{uu}^{(2,1)}$ and $\mathcal{G}_{uu}^{(2,2)}$ with $u\in x,y$. These corrections modify the effective coupling strength as $J\rightarrow J+\delta J$, where
\begin{eqnarray}
\delta J&=&\frac{J}{2}(\xi_1+\xi_2)T\, {\rm Im}(\mathcal{G}_{uu}^{(1)})/{\rm Im}(\mathcal{G}_{uu}^{(0)}) \nonumber \\&-& \frac{J}{4}(\xi^2_1+\xi^2_2)T^2 \,{\rm Im}(\mathcal{G}_{uu}^{(2,1)})/{\rm Im}(\mathcal{G}_{uu}^{(0)})  \\&-&  \frac{J}{4}(\xi_1\xi_2)T^2 \,{\rm Im}(\mathcal{G}_{uu}^{(2,2)})/{\rm Im}(\mathcal{G}_{uu}^{(0)}).\nonumber
\end{eqnarray}
For a single noise realization the corresponding error for the entanglement fidelity is then given by $\mathcal{E}_{\rm e}=\pi^2/4^2 (\delta J/J)^2$. In Table~\ref{table:2} we show the values of the relevant parameters for different $m$. Note that, while the value of ${\rm Im}(\mathcal{G}_{uu}^{(0)})\times (2\pi m)$ is constant, the other parameters depend on $m$. For example, parameter ${\rm Im}(\mathcal{G}_{uu}^{(1)})$ is zero for $m=4,12,20,\dots$, but non-zero for $m=8,16,\dots$. For $m\gg1$, all these parameters converge to the values ${\rm Im}(\mathcal{G}_{uu}^{(1)})/{\rm Im}(\mathcal{G}_{uu}^{(0)}) \simeq 0$, ${\rm Im}(\mathcal{G}_{uu}^{(2,1)})/{\rm Im}(\mathcal{G}_{uu}^{(0)}) \simeq 0.0025$ and ${\rm Im}(\mathcal{G}_{uu}^{(2,2)})/{\rm Im}(\mathcal{G}_{uu}^{(0)}) \simeq -0.005$. In this limit, the expression for $\delta J$ simplifies to
\begin{eqnarray}
\delta J\approx -\frac{JT^2}{4^35^2}(\xi_1-\xi_2)^2,
\end{eqnarray}
and the average error $\mathcal{E}_{\rm e}$ for uncorrelated noise, where $\langle \xi^2_i \xi^2_j \rangle=(2\delta_{ij}+1)\sigma^4$, is given by the result in Eq.~\eqref{eq:entanglement_error}.

\section{Cavity decay}\label{app:CavityDecay}
In this appendix we outline the derivation of an effective master equation for the reduced state of the TLSs, $\mu(t)={\rm Tr}_c\{\rho(t)\}$, where $\rho(t)$ is the full density operator obeying the master equation in Eq.~\eqref{eq:fullME}.
To do so we start with the JC Hamiltonian in the toggling frame, which we write as
\begin{equation}
\tilde H(t) = \Sigma(t) a^\dag(t) + \Sigma^\dagger(t) a(t).
\end{equation} 
Here, $a(t)$ is the cavity mode operator in the interaction picture and   
\begin{equation}
\Sigma(t)=  \sum_{j} \frac{g_j}{2} \left[ f_x(t)  \sigma_j^x - i f_y(t)  \sigma_j^y\right],
\end{equation}
assuming ideal $\pi$-pulses for simplicity. Under the validity of the usual Born-Markov approximation, we can then follow the standard steps for the derivation of a master equation~\cite{QuantumNoise} and we obtain
\begin{equation} 
\dot \mu(t) = - \int_0^t dt' \langle a(t)a^\dag (t') \rangle_c  [\Sigma^\dag(t) , \Sigma(t') \mu(t) ]  +{\rm H.c.}, 
\end{equation} 
where  $\langle a(t) a^\dag(t')\rangle_c= e^{-(i\Delta+\kappa/2)(t-t')}$. In view of the periodicity of the DD sequence, it is more relevant to evaluate the change of the density operator over one period $T$. Therefore, we introduce the discrete time derivative 
\begin{equation} 
\begin{split} 
\left.\frac{\Delta \mu}{\Delta t}\right|_{t=nT} :=  \frac{\mu(nT+T)-\mu(nT)}{T}
\end{split} 
\end{equation}
and take the formal limit $T\rightarrow0$ and $\Delta \mu/\Delta t\rightarrow \dot \mu$ afterwards. This leaves us with the coarse-grained master equation 
\begin{equation} \label{eq:effective_ME}
\begin{split} 
\dot \mu = &- \sum_{i,j}\sum_{\{u,v\}=\{x,y\}}  \mathcal{K}_{uv}^{ij}   [\sigma^{u}_i , \sigma_j^{v} \mu(t) ]  +{\rm H.c.}\\
\equiv &-i \left(H_{\rm nh} \mu - \mu H_{\rm nh}^\dag\right)+ \mathcal{J}_{\rm rec}(\mu ).
\end{split} 
\end{equation} 
Here we have introduced the complex quantities 
\begin{equation}\label{eq:Kijuv}
\begin{split} 
\mathcal{K}_{uv}^{ij}   = \lim_{n\rightarrow \infty} \frac{g_i g_j}{4T}  & \int_{nT}^{(n+1)T} dt \int_0^t dt'  \,  \\
&\times  e^{-(i\Delta+\kappa/2)(t-t')} \tilde f_{u}^*(t) \tilde f_{v} (t'),
\end{split} 
\end{equation} 
where $\tilde{f}_{x}(t)={f}_{x}(t)$, $\tilde{f}_{y}(t)=-i{f}_{y}(t)$. 
By neglecting the recycling term $\mathcal{J}_{\rm rec}(\mu )$, the master equation in Eq.~\eqref{eq:effective_ME} is equivalent to the evolution of the TLSs under the non-Hermitian Hamiltonian 
\begin{equation}
H_{\rm nh} =  - i \sum_{i,j}\sum_{\{u,v\}=\{x,y\}}  \ \mathcal{K}_{uv}^{ij}\sigma^{u}_i  \sigma_j^{v}. 
\end{equation}  
Thus, the real parts of the $\mathcal{K}_{uv}^{ij} $ correspond to effective loss rates, which reduce the norm of the wavefunction. 

To evaluate the remaining integrals, we first calculate the integral over $t'$. To do so we write $\Delta=2\pi m/T +\Delta_{\rm eff}$ and split the integral into a part up to time $nT$ and the rest. For large enough $n$, such that $nT\kappa \gg 1$, we obtain
\begin{equation}
\begin{split} 
&\int_0^t dt'  \,  e^{-(i\Delta+\kappa/2)(t-t')} \tilde f_{v} (t') = \int_{nT}^{t} dt' \, \tilde f_{v} (t') e^{-(i\Delta+\kappa/2)(t-t')}\\
&+ \left( \frac{e^{-(i\Delta+\kappa/2)(t-nT)} }{1-e^{-(i\Delta_{\rm eff}+\kappa/2)T}}\right)\int_0^T dt' \, \tilde f_v(t')e^{-(i\Delta +\kappa/2)(T-t')}.
\end{split} 
\end{equation}
After reinserting this result back into Eq.~\eqref{eq:Kijuv} and combining all the terms, we end up with 
\begin{equation}\label{eq:KxyResult} 
\begin{split} 
\mathcal{K}_{uv}^{ij}    =  \frac{g_i g_j}{4}  \left( \frac{T \tilde{\mathcal{J}}_{uv}(\kappa)}{1-e^{-(i\Delta_{\rm eff}+\kappa/2)T}}\right),
\end{split} 
\end{equation} 
where 
\begin{equation} 
\begin{split} 
 \tilde{\mathcal{J}}_{uv}(\kappa) =  \int_0^T \frac{dt}{T} \int_{t-T}^{t} \frac{dt'}{T}  \, e^{-(i\Delta+\kappa/2)(t-t')} \tilde f^*_{u}(t)  \tilde f_{v} (t').
\end{split} 
\end{equation}

\subsubsection*{Discussion}

For a single spin the decay part of the non-Hermitian Hamiltonian, $H_{\rm decay}=(H_{\rm nh}-H_{\rm nh}^\dag)/2$, reduces to
\begin{equation}
H_{\rm decay} = - \frac{i}{2}\left(  \gamma^e_{\rm eff} |e\rangle\langle e|+ \gamma^g_{\rm eff}  |g\rangle\langle g | \right),
\end{equation}
where we  have introduced the effective decay rates 
\begin{equation}
  \gamma^{e/g}_{\rm eff} = 2{\rm Re}\left\{ \mathcal{K}_{xx} + \mathcal{K}_{yy} \pm i (\mathcal{K}_{xy}-\mathcal{K}_{yx})\right\}.
\end{equation}
To obtain a more meaningful expression, we consider the limit $|i\Delta_{\rm eff}+\kappa/2|T\ll1$, in which case we can approximate $\tilde{\mathcal{J}}_{uv}(0) \simeq   \tilde \eta_u^* \tilde \eta_\nu$, where $\tilde \eta_x=\eta_x$ and $\tilde \eta_y=-i\eta_y$.
Therefore, for a pulse sequence with a non-vanishing first-order coupling, the main contribution to the decay rate is approximately given by
\begin{equation}\label{eq:ResonantDecay}
\begin{split}
 \gamma_{\rm eff}^{e/g} \approx \,\frac{g^2|\eta_x\pm \eta_y|^2\kappa }{4\Delta_{\rm eff}^2+\kappa^2}.
 \end{split} 
\end{equation}
For example, for the case of an effective JC evolution with $\eta_x=\eta_y=\eta$, we recover the result $\gamma_{\rm eff}^{g}\approx 0$ and $\gamma_{\rm eff}^{e}\approx g_{\rm eff}^2\kappa/(\Delta_{\rm eff}^2+\kappa^2/4)$, as expected for a weakly coupled JC model.

When the first-order coupling vanishes, i.e., $\eta_{x,y}=0$, the corrections to $\tilde{\mathcal{J}}_{uv}(0)$ for finite $\kappa$ must be taken into account. In this situation it is more convenient to express the $\mathcal{K}_{uv}^{ij}$ in terms of a Fourier series
\begin{equation}\label{eq:KxyFourier} 
\begin{split} 
\mathcal{K}_{uv}^{ij}    =  \frac{g_i g_j}{4}   \sum_{n=-\infty}^\infty \frac{\left[\tilde\eta_n^{(u)}\right]^*\tilde\eta_n^{(v)}}{i(\Delta-2\pi n/T)+\kappa/2},
\end{split} 
\end{equation} 
where
\begin{equation}
\tilde\eta_n^{(u)}= \frac{1}{T}\int_0^T dt \,  \tilde f_u(t) e^{i2\pi n t/T}.    
\end{equation}
For the case of a single TLS we obtain 
\begin{equation}
\begin{split}
\gamma_{\rm eff}^{e/g}= \,\frac{g^2 \kappa}{4}\sum_{n=-\infty}^\infty \frac{|\eta_n^{\pm}|^2}{(\Delta-2\pi n/T)^2+\kappa^2/4},
 \end{split} 
\end{equation}
where
\begin{equation}
\eta_n^{\pm}= \frac{1}{T}\int_0^T dt \, \left[f_x(t) \pm f_y(t)\right] e^{i2\pi n t/T}.    
\end{equation}
From this expression we immediately see that in the limit of very large detuning, $\Delta T \rightarrow \infty$, the two rates for the ground and the excited state are approximately the same and given by
\begin{equation}\label{eq:decayeg}
\begin{split}
\gamma_{\rm eff}^{e/g}\approx \,&  \frac{g^2}{4\Delta^2} \kappa \sum_n |\eta_n^{\pm}|^2 \\
 =\, & \frac{\gamma_0}{4} \frac{1}{T}\int_0^T dt\, \left[f_x(t) \pm f_y(t)\right]^2  = \frac{\gamma_0}{2},\\
 \end{split} 
\end{equation}
with $\gamma_0=(g/\Delta)^2\kappa$. The last equality in Eq.~\eqref{eq:decayeg} holds for sequences with $\int_0^T dt\, f_x(t)f_y(t)=0 $, which is fulfilled for all noise-canceling sequences. Note that for the undriven case $f_x(t)=f_y(t)=1$ and we recover the usual result $\gamma_{\rm eff}^{g}=0$ and $\gamma_{\rm eff}^{e}\approx \gamma_0=g^2\kappa/\Delta^2$. 

\begin{figure}
	\includegraphics[width=1\columnwidth]{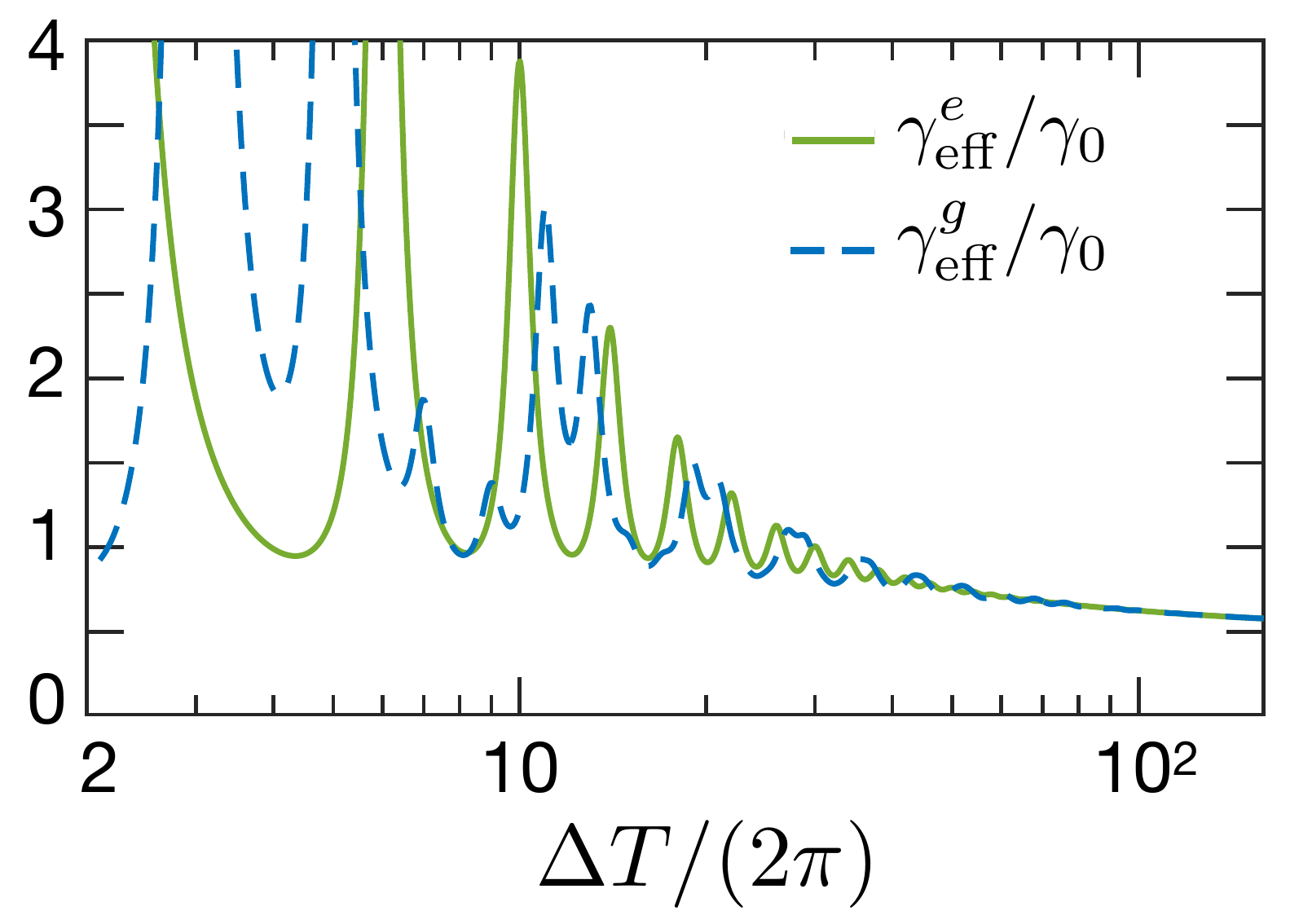}
	\caption{Plot of the effective decay rates $\gamma_{\rm eff}^{e/g}$ as a function of the cavity detuning $\Delta$ for a single TLS driven by a XY8 sequence with $t_1=\tau/2$ and $\tau_\pi=10^{-2}\tau$. }
	\label{Fig:App2}
\end{figure}

As illustrated in Fig.~\ref{Fig:App2} for the XY8 sequence, the dependence of the effective decay rates for intermediate values of $\Delta$ is more involved and exhibits multiple resonances. These resonances occur whenever $\Delta\approx 2\pi m /T$ and $\eta_m^{\pm}\neq 0$, in which case $\gamma_{\rm eff}^{e/g}$ will again be approximately given by Eq.~\eqref{eq:ResonantDecay} (note that $\eta_x\pm\eta_y=\eta_{m}^{\pm}$). On the contrary, if around the $m$-th harmonic, $\eta_m^{\pm}= 0$, then $\gamma_{\rm eff}^{e/g}$ will be determined by the residual sum
\begin{equation}
\begin{split}
\gamma_{\rm eff}^{e/g}\approx \,\frac{g^2 \kappa}{4\Delta^2}\sum_{n\neq m} \frac{|\eta_n^{\pm}|^2}{ (1-n/m)^2},
 \end{split} 
\end{equation}
which, up to numerical prefactors, is comparable to $\gamma_0$.

Note that in the case of $N=2$ and $g_1\simeq g_2$ also the fact that both TLSs couple to the cavity mode symmetrically must be taken into account. In the far-detuned regime, the effective decay is then determined by the non-Hermitian Hamiltonian
\begin{equation}
H_{\rm decay}= -i\frac{g^2\kappa}{2\Delta^2} \left( \mathbbm{1}+ \sigma_1^+\sigma_2^-+\sigma_1^-\sigma_2^+\right). 
\end{equation}
Within the subspace $\{|eg\rangle,|ge\rangle\}$, this leads to the same decay dynamic as for the case of two undriven TLSs. Therefore, also the same effective decay rates appear in the expressions for the entanglement fidelities in Eq.~\eqref{eq:ErrorModelFree} and Eq.~\eqref{eq:ErrorModelXY8}.

\section{Perturbation theory for the JC state transfer}\label{app:PerThe}
To calculate the transfer error for the JC Hamiltonian, we first identify the effective Hamiltonian term responsible for the error $\tilde H_{\rm err}$, with coupling $\epsilon$. The total effective Hamiltonian is then $\tilde H= \tilde H_{\rm int}^{\rm eff}+\tilde H_{\rm err}$, where $\tilde H_{\rm int}^{\rm eff} = \eta g \left(\sigma_-  a^\dagger e^{i\phi}  +{\rm H.c.} \right)$ is the ideal effective Hamiltonian. For small errors $\epsilon T_{\rm t}\ll1$, the effective evolution operator $e^{-i\tilde H T_{\rm f}}$ can be approximated using the Dyson series as
\begin{equation}
   U\approx U_{\rm t} \Big\{1-i \int  \tilde V(t) -\iint \tilde V(t)\tilde V(s)\Big\},
\end{equation}
where $U_{\rm t}=e^{-i\tilde H_{\rm int}^{\rm eff} T_{\rm t}}$, $\tilde V(t)=e^{i\tilde H_{\rm int}^{\rm eff} t}\tilde H_{\rm err}e^{-i\tilde H_{\rm int}^{\rm eff} t} $ and $\int\equiv\int_0^{T_{\rm f}} dt $ and $\iint\equiv\int_0^{T_{\rm f}}dt\int_0^t ds $ for clarity. The state fidelity between the ideal final state $U_{\rm t}|\psi_0\rangle$ and the perturbed final state $U|\psi_0\rangle$ is then $\mathcal{F}=1-\mathcal{E}$, where
\begin{equation}
   \mathcal{E}\approx 2 \iint  \langle\psi_0| \tilde V(t)\tilde V(s)|\psi_0\rangle- \left[\int  \langle\psi_0| \tilde V(t)|\psi_0\rangle\right]^2.
\end{equation}

\end{document}